\documentclass{jpp}
\usepackage{graphicx}

\usepackage[utf8]{inputenc}
\usepackage[T1]{fontenc}
\usepackage{amsmath}
\usepackage{hyperref}
\usepackage{subfig}

\begin{document}

\title{Investigation of turbulent transport regimes in the tokamak edge by using two-fluid simulations}

\author{M. Giacomin\aff{1}
  \corresp{\email{maurizio.giacomin@epfl.ch}}
  and P. Ricci\aff{1}}

\affiliation{\aff{1}Ecole Polytechnique Fédérale de Lausanne (EPFL), Swiss Plasma Center (SPC), CH-1015 Lausanne, Switzerland }

%\date{\today}

\maketitle

\begin{abstract}

The results of flux-driven, two-fluid simulations in single-null configurations are used to investigate the processes determining the turbulent transport in the tokamak edge. Three turbulent transport regimes are identified: (i) a developed transport regime with turbulence driven by an interchange instability, which shares a number of features with the standard L-mode of tokamak operation, (ii) a suppressed transport regime, characterized by a higher value of the energy confinement time, low-amplitude relative fluctuations driven by a Kelvin-Helmholtz instability, a strong $\mathbf{E}\times\mathbf{B}$ sheared flow, and the formation of a transport barrier, which recalls the H-mode, and (iii) a degraded confinement regime, characterized by a catastrophically large interchange-driven turbulent transport, which reminds the crossing of the Greenwald density limit.
We derive an analytical expression of the pressure gradient length in the three regimes.
The transition from the developed to the suppressed transport regime is obtained by increasing the heat source or decreasing the collisionality and vice versa for the transition from the developed transport regime to the degraded confinement regime. 
An analytical expression of the power threshold to access the suppressed transport regime, linked to the power threshold for H-mode access, as well as the maximum density achievable before entering the degraded confinement regime, related to the Greenwald density, are also derived.
The experimental dependencies of the power threshold for H-mode access on density, tokamak major radius, and isotope mass are retrieved. The analytical estimate of the density limit contains the correct dependence on the plasma current and on the tokamak minor radius.

\end{abstract}

\section{Introduction}

The turbulent plasma dynamics in the edge plays a key role in determining the overall performances of a tokamak by governing its confinement properties. Indeed, fundamental phenomena, such as the L-H transition \citep{Wagner1982} and the density limit \citep{greenwald1988,greenwald2002}, strongly depend on the plasma dynamics in the tokamak edge.
Because of the persisting uncertainties in the fundamental understanding of these phenomena, the design of future fusion devices is based on scaling laws.

A scaling law for the power threshold for the L-H transition, $P_\text{\tiny{LH}}$, has been proposed by~\citet{Martin2008a} based on an international H-mode threshold power database:
\begin{equation}
    \label{eqn:power_experiments}
    P_{\text{\tiny{LH}}} \propto n_e^{0.78\pm 0.04} B_T^{0.77\pm 0.03}a^{0.98\pm 0.08}R^{1.0\pm 0.1},
\end{equation}
where $n_e$ is the line-averaged electron density, $B_T$ is the toroidal magnetic field at the tokamak axis, $a$ is the tokamak minor radius and $R$ is the tokamak major radius. In addition, it has been experimentally observed that $P_\text{\tiny{LH}}$ in a single-null geometry is lower when the ion-$\nabla B$ drift direction is towards the X-point, rather than away from it \citep{asdex1989} and that $P_\text{\tiny{LH}}$ depends inversely on $m_i/m_e$  \citep{righi1999,maggi2017}.
Experimental observations in Alcator C-Mod \citep{Snipes1996} and DIII-D \citep{Thomas1998} tokamaks have pointed out the presence of hysteresis in the L-H transition, although this is not a feature universally observed \citep{ryter2013}.
Furthermore, just before the L-H transition, it has been experimentally observed the formation at the tokamak edge of a clear well in the radial electric field profile that induces a strong $\mathbf{E}\times\mathbf{B}$ shear flow, which, in turn, suppresses plasma turbulence \citep{groebner1990,Burrell1997,ryter2015}.
While several models have attempted to uncover the mechanism behind the L-H transition, there is no theory that accounts for all the observations \citep{connor2000}.

The density limit represents the maximum plasma density achievable in tokamaks before the plasma develops a strong MHD activity that leads to the degradation of particle confinement or even a disruption. An experimental scaling law for the density limit, denoted as Greenwald density $n_G$, has been derived by~\citet{greenwald1988},
\begin{equation}
\label{eqn:greenwald}
    n_G = \frac{I_p}{\pi a^2},
\end{equation}
where $I_p$ is the plasma current in MA, $a$ is the tokamak minor radius in m, and $n_G$ is the line-averaged density in $10^{20}$~m$^{-3}$.
Experimental observations show that the cooling of the plasma edge is a key element that characterizes the density limit \citep{vershkov1974,fielding1977}.
In fact, experimental studies reveal that the density limit  can  be  exceeded by  operating  with  peaked density  profiles \citep{kamada1991,lang2012,mahdavi2002,valovic2002},  thus providing a strong evidence of the link between the density limit and edge physics \citep{greenwald2002}.
It has been experimentally observed by \citet{hong2017} that, when the line-averaged density approaches the density limit,  the edge shear flow collapses and, consequently, the turbulent transport strongly increases near the separatrix.    
While there is no widely accepted first-principles model for the density limit, research in this area has focused on mechanisms which lead to strong edge cooling, in particular on the effect of the plasma collisionality on enhanced turbulent transport \citep{greenwald2002}.

The first attempts to provide a unified theoretical description of turbulent transport in the tokamak edge that includes the L-mode confinement regime, the H-mode confinement regime, and a degraded confinement regime, related to the crossing of the density limit, are discussed by \citet{Scott1997} and \citet{rogers1998} in a circular and sheared geometry, based on fluid flux-tube simulations. The transitions from the L-mode to the H-mode and from the L-mode to the density limit are observed by changing the value of the plasma collisionality and $\beta$. The dependence of edge transport on these parameters was then experimentally observed by~\citet{labombard2005}. 
A more recent work \citep{hajjar2018} based on the Hasegawa-Wakatani model \citep{hasegawa1983} in the low $\beta$ limit shows that both the dynamics that characterizes the L-H transition and the density limit can be described as the result of varying the plasma collisionality. By changing the collisionality, three different regimes are identified: a low confinement regime, a high confinement regime, and a regime of degraded particle confinement, which is associated to the density limit. 

The goal of the present manuscript is to extend previous investigations of the edge turbulent regimes by considering a more realistic geometry, i.e. a lower single-null configuration, while retaining the coupling between the edge and both the core and the scrape-off layer (SOL), as a crucial element in determining the plasma dynamics at the tokamak edge. 
In fact, the transport mechanisms occurring in the tokamak periphery are expected to result from a complex interplay among core, edge, and SOL physics  \citep{Fichtmuller1998,Pradalier2017,grenfell2019}, which is difficult to properly model with a simulation domain that does not include all of them. 
As a consequence, we perform turbulence simulations of the whole tokamak in order to approach this interplay. 

Turbulence in the tokamak core is most often simulated by means of gyrokinetic codes, while fluid codes are usually applied in the SOL, taking advantages of its higher plasma collisionality. This separation undermines the possibilities to advance our understanding of the plasma dynamics in the tokamak edge.
For this reason, recently, significant effort was carried out in order to extend gyrokinetic models towards the edge and the SOL \citep{qin2007,hahm2009,frei2019}. The first gyrokinetic simulation of the L-H transition that encompasses the edge and the SOL was carried out by using the XGC1 code \citep{chang2017,ku2018}.  
Since the computational cost of a gyrokinetic simulation of the L-H transition on a global transport time-scale remains prohibitively high \citep{chang2017}, an ion heat flux at the edge was imposed in the XGC1 L-H simulation, considerably larger than the experimental one.
This large flux allowed a reduced computational cost of the simulation, as the L-H transition was due to fast electrostatic bifurcation occurring on time scale considerably shorter than the one required to reach the global steady state transport conditions. 
Other efforts to extend gyrokinetic codes to simulate turbulence in open-field-line systems include the Gkeyll \citep{Shi2017}, GENE \citep{Pan2018}, ELMFIRE \citep{chone2018}, and COGENT \citep{dorf2020}  codes.
In this paper, we follow a different approach and we extend fluid simulations to the core region, in order to cover the whole tokamak plasma volume. While not providing an accurate description of turbulence in the core, these simulations allow us to explore the parameter space of edge turbulence at different values of heat source and plasma collisionality in a global transport steady state that is the result of the heat and particle sources in the core, turbulent transport, and the losses at the vessel. 
Thanks to these simulations, we draw a portrait of the edge turbulent regimes that can be used as a basis to interpret the results of more complete kinetic simulations. 

Our study is based on simulations carried out with GBS \citep{Ricci2012,halpern2016,Paruta2018},  a three-dimensional, flux-driven, two-fluid simulation code that has been developed to study plasma turbulence in the tokamak boundary. Similarly to other turbulent codes, such as BOUT++ \citep{Dudson2015}, GDB \citep{Zhu2018}, GRILLIX \citep{Stegmeir2018a}, HESEL \citep{Nielsen2015}, and TOKAM3X \citep{Tamain2016}, GBS evolves the drift-reduced Braginskii's equations \citep{Zeiler1997}, a set of two-fluid equations valid to describe phenomena occurring on time scales longer than $1/\Omega_{ci}$, with $\Omega_{ci}=eB_T/m_i$ the ion cyclotron frequency, perpendicular length scales longer than the ion Larmor radius, and parallel length scales longer than the mean free path.  

Early fluid simulations performed with the BOUT code have already shown that the physics of the L-H transition can be addressed by means of fluid models \citep{xu2000,xu2002}, even though fluid models exclude a large fraction of modes that are relevant to edge transport, e.g. trapped electron modes, electron temperature gradients, microtearing modes, and kinetic ballooning modes, while retaining the fluid limit of the ion temperature gradient modes \citep{Mosetto2015}. Later numerical investigations of the L-H transition have been carried out by using 2D and 3D fluid simulations and have pointed out the spontaneous formation of a transport barrier \citep{rasmussen2015,chone2014,chone2015}.    
Indeed, despite their simplicity, our simulations show the presence of three turbulent transport regimes: (i) a developed transport regime, which we associate to the standard L-mode, (ii) a suppressed transport
regime, characterized by a higher value of the energy confinement time due to the onset of a transport barrier near the separatrix, and a lower relative fluctuation level, with features that recall the H-mode, and (iii) a degraded confinement regime, characterized by a catastrophically  large  turbulent  transport, which we  link  to  the  density  limit.
In the developed transport regime and degraded confinement regime, turbulent transport is driven by the interchange instability, while in the suppressed transport regime by the Kelvin-Helmholtz (KH) instability. 
We then analyse the transitions between these regimes. As the heat source increases, a transition from the developed transport regime to the suppressed transport regime is observed. This transition is due to the formation of a strong $\mathbf{E}\times\mathbf{B}$ shear across the separatrix, which stabilizes the interchange instability and destabilizes the KH instability. At the transition, a transport barrier forms at the tokamak edge and, consequently, the energy confinement time increases by approximately a factor of two. In fact, the transition from the developed transport regime to the suppressed transport regime shows common features to the L-H transition observed in the experiments.
By imposing a flux balance at the separatrix between perpendicular and parallel transport, we then derive an equation for the heat source threshold, which can be identified as the power threshold for H-mode access, that we compare to the experimental scaling law of Eq.~\eqref{eqn:power_experiments}. 
The transition from the developed transport regime to the degraded confinement regime is obtained by increasing the normalized plasma collisionality, proportional to the plasma density, or by reducing the heat source. We derive an analytical estimate of the maximum density achievable before accessing to the degraded confinement regime. The estimate is compared to the Greenwald density limit of Eq.~\eqref{eqn:greenwald}.  

The present paper is organized as follows. In~\S\ref{sec:model}, we describe the physical model considered to study turbulent transport in the tokamak edge. An overview of simulation results is presented in~\S\ref{sec:overview}, where we discuss the observation of three turbulent transport regimes. In~\S\ref{sec:transport_regimes}, we derive the analytical expressions of the equilibrium pressure gradient length in the three transport regimes. The heat source threshold to access the suppressed transport regime and the density threshold to access the degraded confinement regime are derived in~\S\ref{sec:transitions}. The conclusions follow in~\S\ref{sec:conclusions}.   

\section{Simulation model\label{sec:model}}

Our investigations are based on a drift-reduced Braginskii two-fluid plasma model implemented in the GBS code \citep{Ricci2012,halpern2016,Paruta2018}. 
The application of drift-reduced fluid models to the study of plasma turbulence is valid when the electron mean-free path is shorter than the parallel connection length, $\lambda_e \ll L_\parallel \sim 2\pi q R$, and the dominant modes develop on perpendicular scale lengths larger than the ion Larmor radius, $k_\perp \rho_i \ll 1$.
The high collisionality required by fluid models is typically observed in the edge of L-mode discharges. Regarding the H-mode, we note that, for typical values of density and temperature at the top of the pedestal for neutral beam heated discharges of a medium size tokamak such as TCV, $\lambda_e/L_\parallel$ ranges from 0.05 ($T_e\simeq 100$~eV and $n\simeq 5\times 10^{19}$~m$^{-3}$) to 0.4 ($T_e\simeq 200$~eV and $n\simeq 3\times 10^{19}$~m$^{-3}$), depending on the external gas injection rate \citep{sheikh2018}, thus providing a justification to the use of a fluid model. On the other hand, in the case of JET tokamak, typical values of density and temperature at the top of the pedestal \citep{beurskens2011} ($T_e\simeq 900$~eV and $n\simeq 7\times 10^{19}$~m$^{-3}$) lead to $\lambda_e/L_\parallel \simeq 80$. 
Focusing on the drift approximation that, contrary to more advanced fluid models (see e.g. \citet{wiesenberger2019}), does not allow us to describe finite-Larmor radius effects, we observe that the dominant modes in our simulations satisfy $k_\perp \rho_i \ll 1$, consistently with our model hypothesis, although turbulence in the tokamak edge can also be driven by unstable modes with $k_\perp \rho_i \sim 1$ \citep{jenko2001,dickinson2012}.

For the sake of simplicity, we consider a rather simple drift-reduced Braginskii two-fluid model for this first exploration of the parameter space. For instance, we consider the electrostatic limit, even if electromagnetic effects are important for the edge turbulent transport in H-mode (see, e.g.,  \citet{wan2013,doerk2015,kriete2020}), by playing a role in constraining the pedestal height and width (see, e.g. \citet{snyder2004,snyder2009}), and by affecting the SOL dynamics at high $\beta$ (see, e.g., \citet{Halpern2013}). The use of the electrostatic limit is motivated by~\citet{hajjar2018}, which shows that, even in the low-$\beta$ limit, different turbulent regimes can be retrieved by varying the plasma collisionality.
We also use the Boussinesq approximation in the evaluation of the polarization current \citep{Ricci2012,yu2006}.
The effect of the Boussinesq approximation is discussed in~\citet{yu2006} and \citet{bodi2011}, showing that it has a negligible effect on SOL turbulence. 
In the edge, the validity of the Boussinesq approximation is addressed in~\citet{stegmeir2019} and \citet{ross2019} showing that there is no substantial difference in the equilibrium profiles when the Boussinesq approximation is considered.
Although in theoretical \citep{chone2014,chone2015} and experimental \citep{viezzer2013} works it is shown that neoclassical corrections can play an important role in the onset of transport barriers and, consequently, in the L-H transition, we do not include these effects in our model. 
Trapped particle modes, which can also play an important role in the L-H transition, especially in low-aspect ratio devices \citep{rewoldt1996,dannert2005}, are neglected here.
Finally, while the neutral dynamics may also have an effect on the L-H transition dynamics, as shown by~\citet{shaing1995,carreras1996,owen1998}, we do not include the interplay between plasma and neutrals, although this is implemented in GBS \citep{wersal2015}. 
Within these approximations, the model equations we consider are the following:
\begin{align}
\label{eqn:density}
\frac{\partial n}{\partial t} =& -\frac{\rho_*^{-1}}{B}\bigl[\phi,n\bigr]+\frac{2}{B}\Bigl[C(p_e)-nC(\phi)\Bigr] 
-\nabla_{\parallel}(n v_{\parallel e}) + D_n\nabla_{\perp}^2 n +s_n\, ,\\
\label{eqn:vorticity}
\frac{\partial \omega}{\partial t} =& -\frac{\rho_*^{-1}}{B}\bigl[\phi,\omega\bigr] - v_{\parallel i}\nabla_\parallel \omega + \frac{B^2}{n}\nabla_{\parallel}j_{\parallel} 
+ \frac{2B}{n}C(p_e + \tau p_i) + \frac{B}{3n}C(G_i) + D_{\omega}\nabla_\perp^2 \omega\, ,\\
\label{eqn:electron_velocity}
\frac{\partial v_{\parallel e}}{\partial t} =& -\frac{\rho_*^{-1}}{B}\bigl[\phi,v_{\parallel e}\bigr] - v_{\parallel e}\nabla_\parallel v_{\parallel e} 
+ \frac{m_i}{m_e}\Bigl(\nu j_\parallel+\nabla_\parallel\phi-\frac{1}{n}\nabla_\parallel p_e-0.71\nabla_\parallel T_e\Bigr) \nonumber \\
&+\frac{4}{3n}\frac{m_i}{m_e}\eta_{0,e}\nabla^2_\parallel v_{\parallel e} + D_{v_{\parallel e}}\nabla_\perp^2 v_{\parallel e}\,, \\
\label{eqn:ion_velocity}
\frac{\partial v_{\parallel i}}{\partial t} =& -\frac{\rho_*^{-1}}{B}\bigl[\phi,v_{\parallel i}\bigr] - v_{\parallel i}\nabla_\parallel v_{\parallel i} - \frac{1}{n}\nabla_\parallel(p_e+\tau p_i)
+ \frac{4}{3n}\eta_{0,i}\nabla^2_\parallel v_{\parallel i} + D_{v_{\parallel i}}\nabla_\perp^2 v_{\parallel i}\, ,\\
\label{eqn:electron_temperature}
\frac{\partial T_e}{\partial t} =& -\frac{\rho_*^{-1}}{B}\bigl[\phi,T_e\bigr] - v_{\parallel e}\nabla_\parallel T_e 
+ \frac{2}{3}T_e\Bigl[0.71\nabla_\parallel v_{\parallel i}-1.71\nabla_\parallel v_{\parallel e}
+0.71 (v_{\parallel i}-v_{\parallel e})\frac{\nabla_\parallel n}{n}\Bigr] \nonumber \\
&+ \frac{4}{3}\frac{T_e}{B}\Bigl[\frac{7}{2}C(T_e)+\frac{T_e}{n}C(n)-C(\phi)\Bigr] 
+ \chi_{\perp e}\nabla_\perp^2 T_e + \chi_{\parallel e}\nabla_\parallel^2 T_e + s_{T_e}\,,\\
\label{eqn:ion_temperature}
\frac{\partial T_i}{\partial t} =& -\frac{\rho_*^{-1}}{B}\bigl[\phi,T_i\bigr] - v_{\parallel i}\nabla_\parallel T_i 
+ \frac{4}{3}\frac{T_i}{B}\Bigl[C(T_e)+\frac{T_e}{n}C(n)-C(\phi)\Bigr] - \frac{10}{3}\tau\frac{T_i}{B}C(T_i) \nonumber \\ 
&+ \frac{2}{3}T_i(v_{\parallel i}-v_{\parallel e})\frac{\nabla_\parallel n}{n} -\frac{2}{3}T_i\nabla_\parallel v_{\parallel e} 
+ \chi_{\perp i}\nabla_\perp^2 T_i + \chi_{\parallel i}\nabla_\parallel^2 T_i + s_{T_i}\,,\\
\label{eqn:poisson}
\nabla_\perp^2\phi =& \omega-\tau\nabla_\perp^2 T_i\ .
\end{align}

In Eqs.~\eqref{eqn:density}-\eqref{eqn:poisson} and in the following of the present paper (unless specified otherwise), the density, $n$, the electron temperature, $T_e$, and the ion temperature, $T_i$, are normalized to the reference values $n_0$, $T_{e0}$, and $T_{i0}$. 
The electron and ion parallel velocities, $v_{\parallel e}$ and $v_{\parallel i}$, are normalized to the reference sound speed $c_{s0}=\sqrt{T_{e0}/m_i}$. The norm of the magnetic field, $B$, is normalized to the reference value $B_T$, which, under the assumption of large aspect ratio \citep{jolliet2014,Paruta2018}, is assumed to be constant. Perpendicular lengths are normalized to the ion sound Larmor radius $\rho_{s0}=c_{s0}/\Omega_{ci}$ and parallel lengths are normalized to the tokamak major radius $R_0$. Time is normalized to $R_0/c_{s0}$. The dimensionless parameters appearing in the model equations are the normalized ion sound Larmor radius, $\rho_* = \rho_{s0}/R_0$, the ion to electron temperature ratio, $\tau = T_{i0}/T_{e0}$, the normalized electron and ion viscosities, $\eta_{0,e}$ and $\eta_{0,i}$, the normalized electron parallel and perpendicular thermal conductivities, $\chi_{\parallel e}$ and $\chi_{\perp e}$, the corresponding ion quantities, $\chi_{\parallel i}$ and $\chi_{\perp i}$, and the normalized Spitzer resistivity, $\nu = e^2n_0R_0/(m_ic_{s0}\sigma_\parallel) = \nu_0 T_e^{-3/2}$, with 
\begin{align}
\sigma_\parallel =& \biggl(1.96\frac{n_0 e^2 \tau_e}{m_e}\biggr)n=\biggl(\frac{5.88}{4\sqrt{2\pi}}\frac{(4\pi\epsilon_0)^2}{e^2}\frac{ T_{e0}^{3/2}}{\lambda\sqrt{m_e}}\biggr)T_e^{3/2},\\
\label{eqn:resistivity}
\nu_0=&\frac{4\sqrt{2\pi}}{5.88}\frac{e^4}{(4\pi\epsilon_0)^2}\frac{\sqrt{m_e}R_0n_0\lambda}{m_i c_{s0} T_{e0}^{3/2}},
\end{align}
where $\lambda$ is the Coulomb logarithm.
We highlight that the normalized Spitzer resistivity depends linearly on the reference density $n_0$.
The numerical diffusion terms, $D_f\nabla_{\perp}^2 f$, are added for numerical stability and they lead to significantly smaller transport than the turbulent processes described by the simulations. By considering typical values at the separatrix of a TCV L-mode discharge (tokamak major radius $R_0 \simeq $ 0.9 m and toroidal magnetic field at the tokamak axis $B_T \simeq$ 1.4 T) as reference density and electron temperature, i.e. $n_0\simeq10^{19}$m$^{-3}$ and $T_{e0} \simeq 20$~eV, we obtain a reference value for the numerical perpendicular diffusion coefficient of the order $10^{-2}$  m$^2$/s, two orders of magnitude smaller than the effective diffusion coefficient due to turbulence. 
The source terms in the density and temperature equations, $s_n$ and $s_T$, are added to fuel and heat the plasma.

The spatial operators appearing in Eqs.~\eqref{eqn:density}-\eqref{eqn:poisson} are the $\mathbf{E}\times\mathbf{B}$ convective term $\bigl[g,f\bigr]=\mathbf{b}\ \cdot\ (\nabla g \times \nabla f)$, the curvature operator $C(f)=B/2\bigl[\nabla \times (\mathbf{b}/B)\bigr]\cdot \nabla f$, the parallel gradient $\nabla_\parallel f=\mathbf{b}\cdot\nabla f$, and the perpendicular Laplacian ${\nabla_\perp^2 f=\nabla\cdot\bigl[(\mathbf{b}\times\nabla f)\times\mathbf{b}\bigr]}$, where $\mathbf{b}=\mathbf{B}/B$ is the unit vector of the magnetic field. 
The toroidally symmetric equilibrium magnetic field is written in terms of the poloidal magnetic flux $\psi$, normalized to $\rho_{s0}^2B_T$, as
\begin{equation}
    \mathbf{B}=\pm\nabla\varphi+\rho_*\nabla\psi\times\nabla\varphi ,
\end{equation}
where $\varphi$ is the toroidal angle, with $\nabla\varphi$ normalized to $R_0$. The plus (minus) sign refers to the direction of the toroidal magnetic field with the ion-$\nabla B$ drift pointing upwards (downwards).
The poloidal magnetic flux is a function of the normalized tokamak major radius $R$ and of the vertical coordinate $Z$, i.e. $\psi=\psi(R,Z)$.
Under the assumption of large aspect ratio, $\epsilon = \frac{a}{R_0} \ll 1$, and poloidal magnetic filed much smaller than the toroidal one, $\delta=\rho_*||\nabla\psi|| \ll 1$, we can compute the differential operators appearing in Eqs.~\eqref{eqn:density}-\eqref{eqn:poisson} by retaining only the zeroth-order terms in $\epsilon$ and $\delta$. In $(R,\varphi,Z)$ toroidal coordinates, the curvature operator in dimensionless units can be expanded as
\begin{equation}
      C(f) = \frac{\rho_*^{-1}}{2 B} \biggl( \frac{B_\varphi}{B^2}\partial_Z B^2 \partial_R f - \frac{B_\varphi}{B^2} \partial_R B^2\partial_Z f \biggr) + O(\epsilon,\delta) \,,
\end{equation}
where $B_\varphi = B_T R_0/R$.
These terms take into account the spatial variation of $B^2$. Since 
\begin{equation}
    B^2 = B_T^2\Bigl(\frac{R_0^2}{R^2} + O(\delta^2)\Bigr)\,,
\end{equation}
its spatial derivatives at zeroth-order in $\epsilon$ and $\delta$ are 
\begin{align}
    \partial_Z B^2 &= 0\\
    \partial_R B^2 &=  -2 \rho_*B_T^2\,.
\end{align}
Finally, the curvature operator at zeroth-order in $\epsilon$ and $\delta$ becomes
\begin{equation}
    C(f) = \pm\partial_Z f + O(\epsilon,\delta)\,. 
\end{equation}
Similar algebra leads to the other differential operators at zeroth-order in $\epsilon$ and $\delta$ (see \citet{Paruta2018} for details). In summary, the differential operators implemented in GBS in $(R,\varphi,Z)$ toroidal coordinates are
\begin{align}
    \label{eqn:poisson_brackets}
    [\phi,f] =& \pm\biggl( \frac{\partial\phi}{\partial Z}\frac{\partial f}{\partial R} - \frac{\partial\phi}{\partial R}\frac{\partial f}{\partial Z} \biggr), \\
    C(f) =& \pm \frac{\partial f}{\partial Z},\\
    \nabla_\parallel f =& \frac{\partial\psi}{\partial Z}\frac{\partial f}{\partial R} - \frac{\partial\psi}{\partial R}\frac{\partial f}{\partial Z}  \pm\frac{\partial f}{\partial \varphi},\\
    \nabla_\perp^2 f =& \frac{\partial^2 f}{\partial R^2} + \frac{\partial^2 f}{\partial Z^2},\\
    \label{eqn:parallel_lapl}
    \nabla_\parallel^2 f =& \nabla_\parallel\bigl(\nabla_\parallel f\bigr)\,,
\end{align}
where the plus (minus) sign is again used for the ion-$\nabla B$ drift pointing upwards (downwards).
For the analysis of the turbulent transport in~\S\ref{sec:transport_regimes}, flux-coordinates ($\nabla\psi,\nabla\chi,\nabla\varphi)$ are considered, where $\nabla\psi$ denotes the direction orthogonal to flux surfaces, $\nabla\varphi$ is the toroidal direction, and $\nabla\chi=\nabla\varphi\times\nabla\psi$. 

\begin{figure}
    \centering
    \includegraphics[scale=0.4]{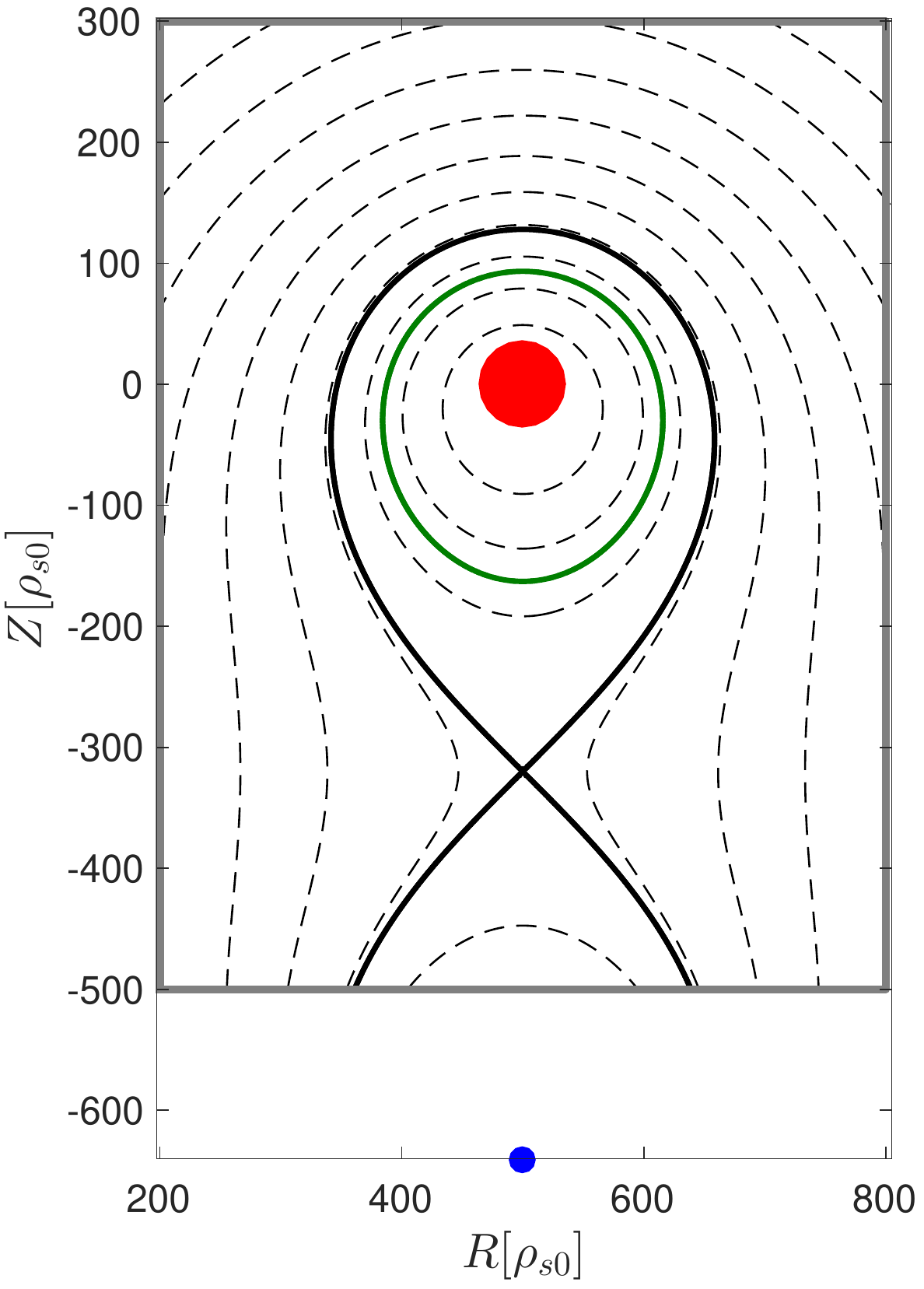}
    \caption{Contour plot of the poloidal flux function $\psi$ considered in the present work (black dashed line). The separatrix is shown as a solid black line. The boundary domain is indicated by a solid grey line. The red circle represents the plasma current, while the blue circle, located outside the domain, represents the current filament used to generate the X-point. The flux surface $\psi = \psi_n =\psi_T$ is shown as a solid green line.}
    \label{fig:equilibrium}
\end{figure}

Similarly to the simulations presented in~\citet{giacomin2020}, we consider Eqs.~\eqref{eqn:density}-\eqref{eqn:poisson} in a rectangular poloidal cross section of size $L_R$ and $L_Z$ in the radial and vertical directions, respectively.
The single-null magnetic configuration used in the simulations presented herein is analytically obtained by solving the Biot-Savart law for a straight current filament, which is located outside the domain, and a current density with Gaussian profile, which is centered at the tokamak magnetic axis, ($R_0$,$Z_0$), and mimics the plasma current (see Fig.~\ref{fig:equilibrium}). The current filament and the plasma current are centered at the same radial position.

The density and the temperature sources are analytical and toroidally uniform functions of $\psi(R,Z)$,  
\begin{align}
    \label{eqn:density_source}
    s_n =& s_{n0} \exp\biggl(-\frac{\bigl(\psi(R,Z)-\psi_{n}\bigr)^2}{\Delta_n^2}\biggr),\\   
    \label{eqn:temperature_source}
    s_T =& \frac{s_{T0}}{2}\biggl[\tanh\biggl(-\frac{\psi(R,Z)-\psi_{T}}{\Delta_T}\biggr)+1\biggr], 
\end{align}
where $\psi_n$ and $\psi_T$, displayed in Fig.~\ref{fig:equilibrium}, are flux surfaces located inside the last closed flux surface (LCFS). The density source is localized around the flux surface $\psi_n$, close to the separatrix, and mimics the ionization process, while the temperature source extends through the entire core and mimics the ohmic heating. 
We define $S_n$ and $S_T$ as the total density and temperature source integrated over the area inside the separatrix,
\begin{equation}
    S_n=\int_{A_{\text{LCFS}}} \rho_* s_n(R,Z)\,\mathrm{d}R\mathrm{d}Z
\end{equation}
and
\begin{equation}
    S_T=\int_{A_{\text{LCFS}}} \rho_* s_T(R,Z)\,\mathrm{d}R\mathrm{d}Z\,,
\end{equation}
where the factor $\rho_*$ appears from the normalization. 
Analogously, we define the electron power source $S_p=\int_{A_{\text{LCFS}}} \rho_* s_p\,\mathrm{d}R\mathrm{d}Z$, with $s_p=n s_{T_e} + T_e s_n$ and $s_{T_e}$ the electron temperature source.

Magnetic pre-sheath boundary conditions, derived by \citet{Loizu2012}, are applied at the target plates. Neglecting correction terms linked to radial derivatives of the density and potential at the target plate, these boundary conditions can be expressed as
\begin{align}
    \label{eqn:boundary_first}
    v_{\parallel i}=& \pm \sqrt{T_e+\tau T_i},\\
    v_{\parallel e}=& \pm \sqrt{T_e+\tau T_i}\  \mathrm{max}\Bigl\{\exp\Bigl(\Lambda-\frac{\phi}{T_e}\Bigr),\exp\bigl(\Lambda\bigr)\Bigr\},\\
    \partial_Z n =& \mp \frac{n}{\sqrt{T_e+\tau T_i}}\partial_Z v_{\parallel i},\\
    \partial_Z \phi =& \mp \frac{T_e}{\sqrt{T_e+\tau T_i}}\partial_Z v_{\parallel i},\\
    \partial_Z T_e =&\ \partial_Z T_i=\ 0,\\
    \label{eqn:boundary_last}
    \omega =& -\frac{T_e}{T_e+\tau T_i}\Bigl[\bigl(\partial_Z v_{\parallel i}\bigr)^2\pm\sqrt{T_e+\tau T_i}\,\partial_{ZZ}^2v_{\parallel i}\Bigr], 
\end{align}
where $\Lambda=3$. The top (bottom) sign refers to the magnetic field pointing towards (away from) the target plate.  

The numerical implementation of Eqs.~\eqref{eqn:density}-\eqref{eqn:poisson} with the boundary conditions given by Eqs.~\eqref{eqn:boundary_first}-\eqref{eqn:boundary_last} in the GBS code is detailed in~\citet{Paruta2018}. 
The differential operators in Eqs.~\eqref{eqn:poisson_brackets}-\eqref{eqn:parallel_lapl} are discretized with a fourth-order finite difference scheme on a non-field aligned grid, which allows for simulations in arbitrary magnetic configurations. 
GBS was verified with the method of manufactured solutions \citep{riva2014}. 
Convergence studies carried out by~\citet{Paruta2018} show that the numerical convergence is retrieved with the considered grid resolution.

\section{Overview of simulation results}\label{sec:overview}

We report on a set of GBS simulations carried out with the following parameters: $\rho_*^{-1}=500$, $a/R_0\simeq 0.3$, $\tau=1$, $\eta_{0,e}=5\times10^{-3}$, $\eta_{0,i}=1$, $L_R=600$, $L_Z=800$, $s_{n0}=0.3$, $\Delta_n = 800$, $\Delta_T = 720$, and $Z_1 = -640~\rho_{s0}$. The parallel and perpendicular thermal conductivities are considered constant parameters, $\chi_{\parallel e}=\chi_{\parallel i}=1$ and $\chi_{\perp e}=\chi_{\perp i}=6$.
We vary $s_{T0}$ and $\nu_0$, considering $s_{T0}=$~\{0.075, 0.15, 0.3, 0.6\}  and $\nu_0=$~\{0.2, 0.6, 0.9, 2.0\}. We consider the same values $s_{T0}$ for both the ion and electron temperature source, although it should be noted that experimental observations \citep{ryter2014,ryter2015} show the importance of the ion heat channel with respect to the electron one in the physics of the L-H transition.
The ion-$\nabla B$ drift direction points upwards (unfavorable for H-mode access) in all the simulations except the ones considered in \S\ref{sec:power_threshold}, where the effect of the toroidal magnetic field direction is discussed.
The value of the plasma current $I_p$ and the width of its Gaussian distribution $\sigma$ are chosen to have the safety factors $q_0\simeq 1$ at the magnetic axis and $q_{95}\simeq 4$. The value of the current in the filament is chosen to be equal to the plasma current.
To connect these parameters to a physical case, we can consider typical values at the separatrix of a TCV L-mode discharge (tokamak major radius $R_0 \simeq $ 0.9 m and toroidal magnetic field at the tokamak axis $B_T \simeq$ 1.4 T) as reference density and electron temperature, i.e. $n_0 \simeq 10^{19}$~m$^{-3}$ and $T_{e0} \simeq 20$~eV, which lead to a size of the simulation domain in physical units $L_R \simeq $ 30 cm, $L_Z \simeq 40$ cm and $R_0 \simeq $ 25 cm, which is approximately 1/3 of the TCV size. 
Regarding the numerical parameters, the grid used is $N_R\times N_Z\times N_\varphi = 240\times 320\times 80$ and the time-step is $2\times 10^{-5}$. 
After an initial transient, the simulations reach a global turbulent quasi-steady state, that results from the interplay between the sources in the closed flux surface region, the turbulence that transports plasma and heat from the core to the SOL, and the losses at the vessel.

\begin{figure}
    \centering
    \subfloat[]{\includegraphics[width=0.39\textwidth]{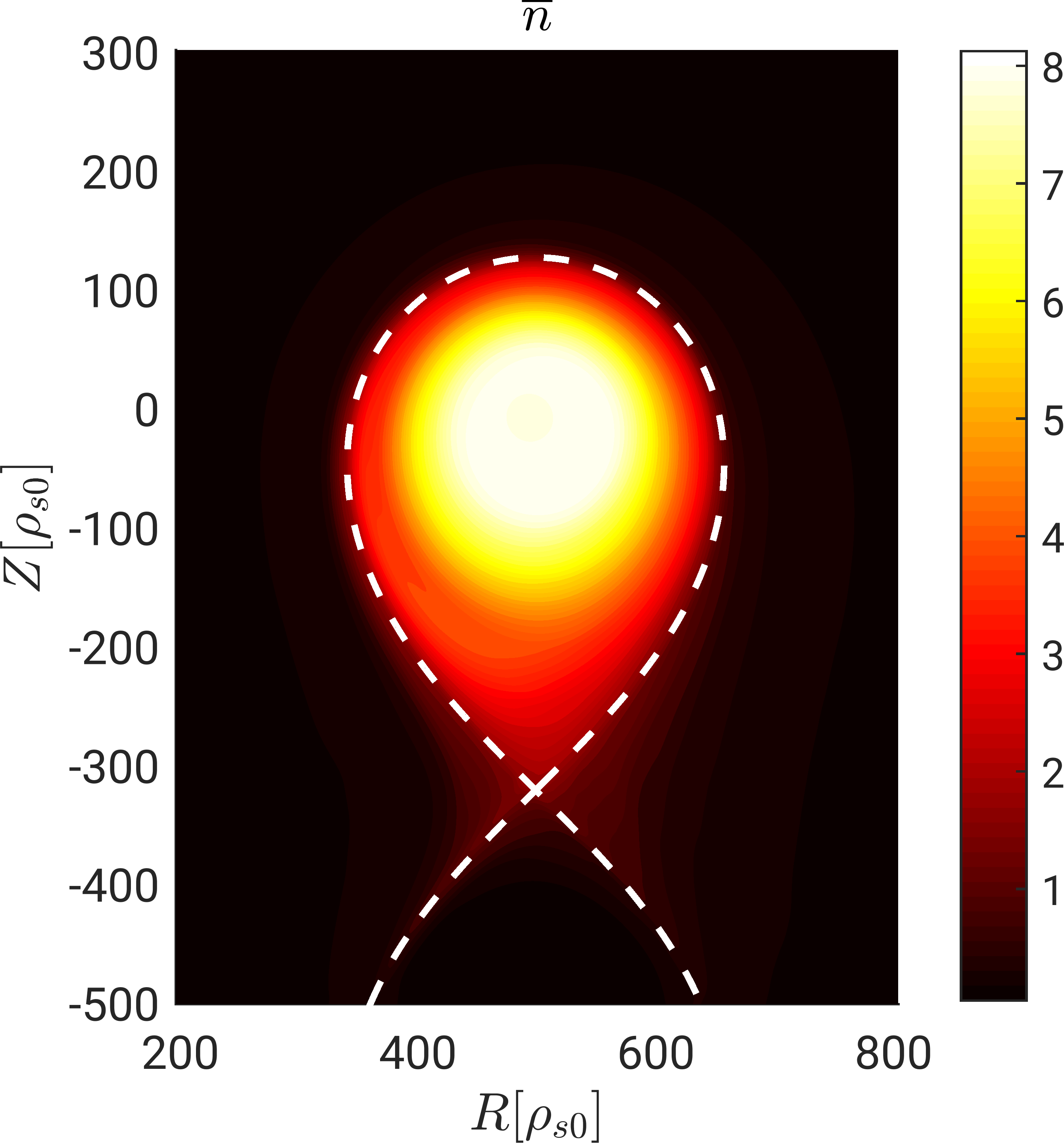}}\quad
    \subfloat[]{\includegraphics[width=0.4\textwidth]{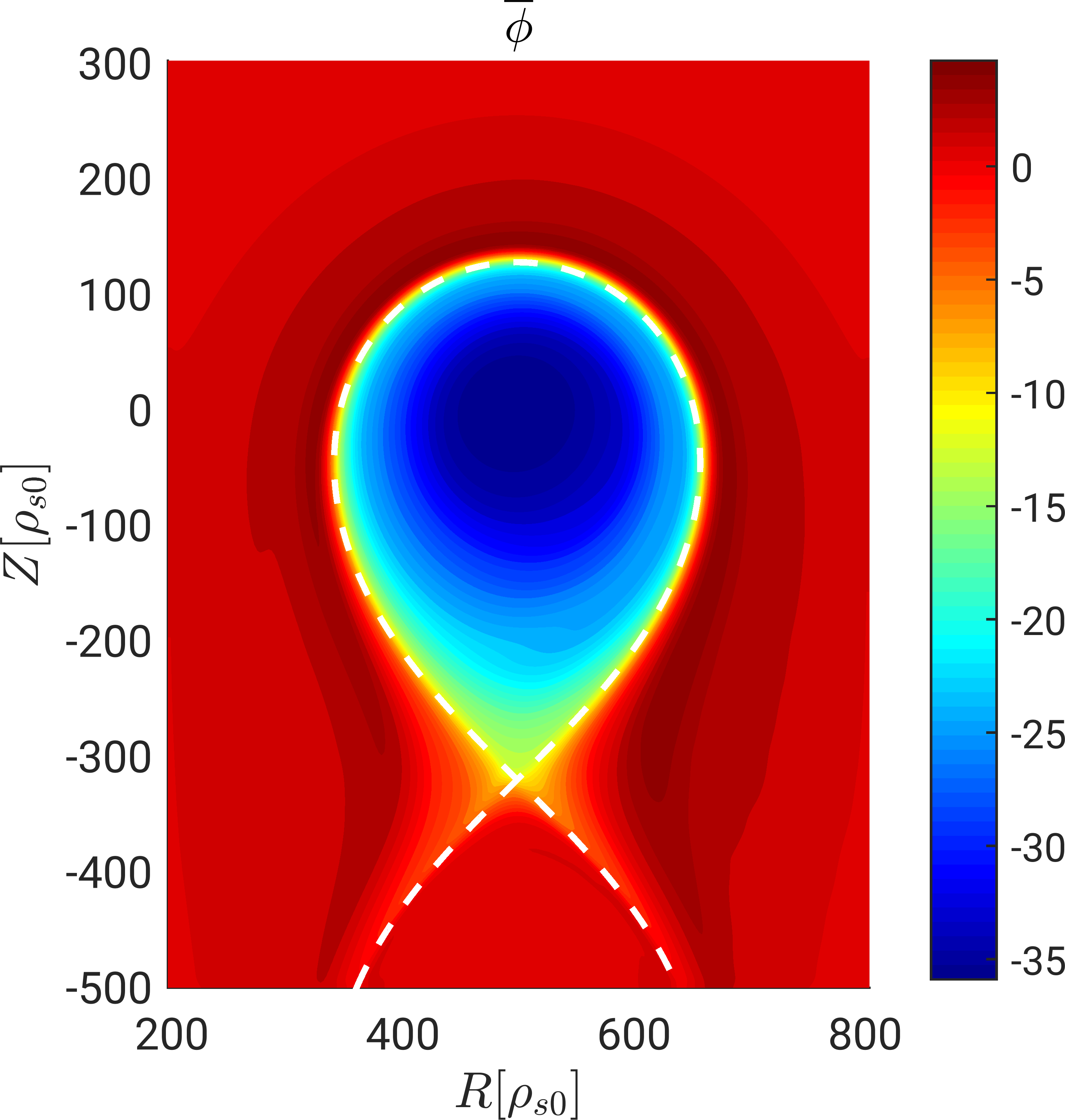}}\\
    \subfloat[]{\includegraphics[width=0.4\textwidth]{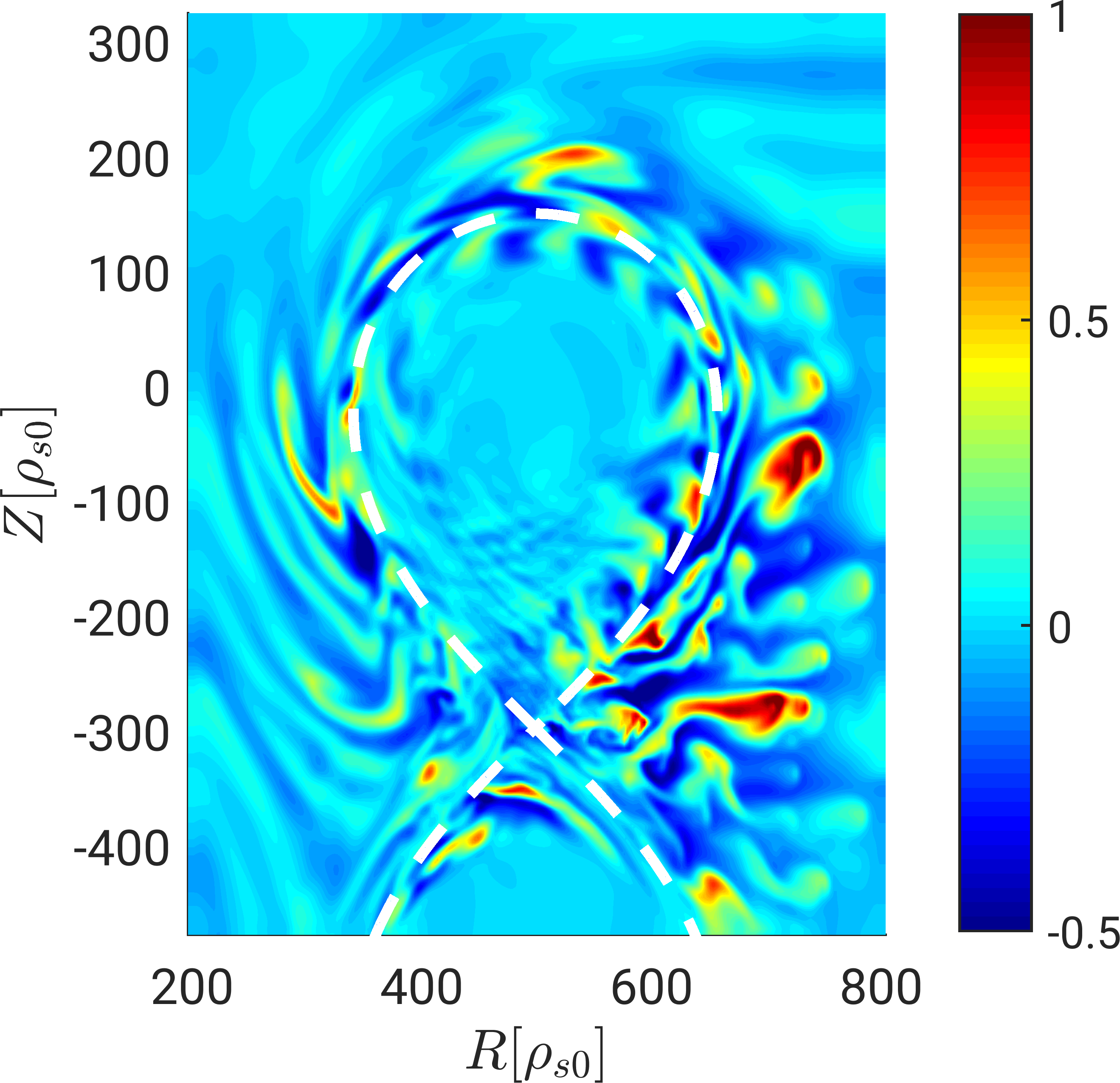}}\quad
    \subfloat[]{\includegraphics[width=0.4\textwidth]{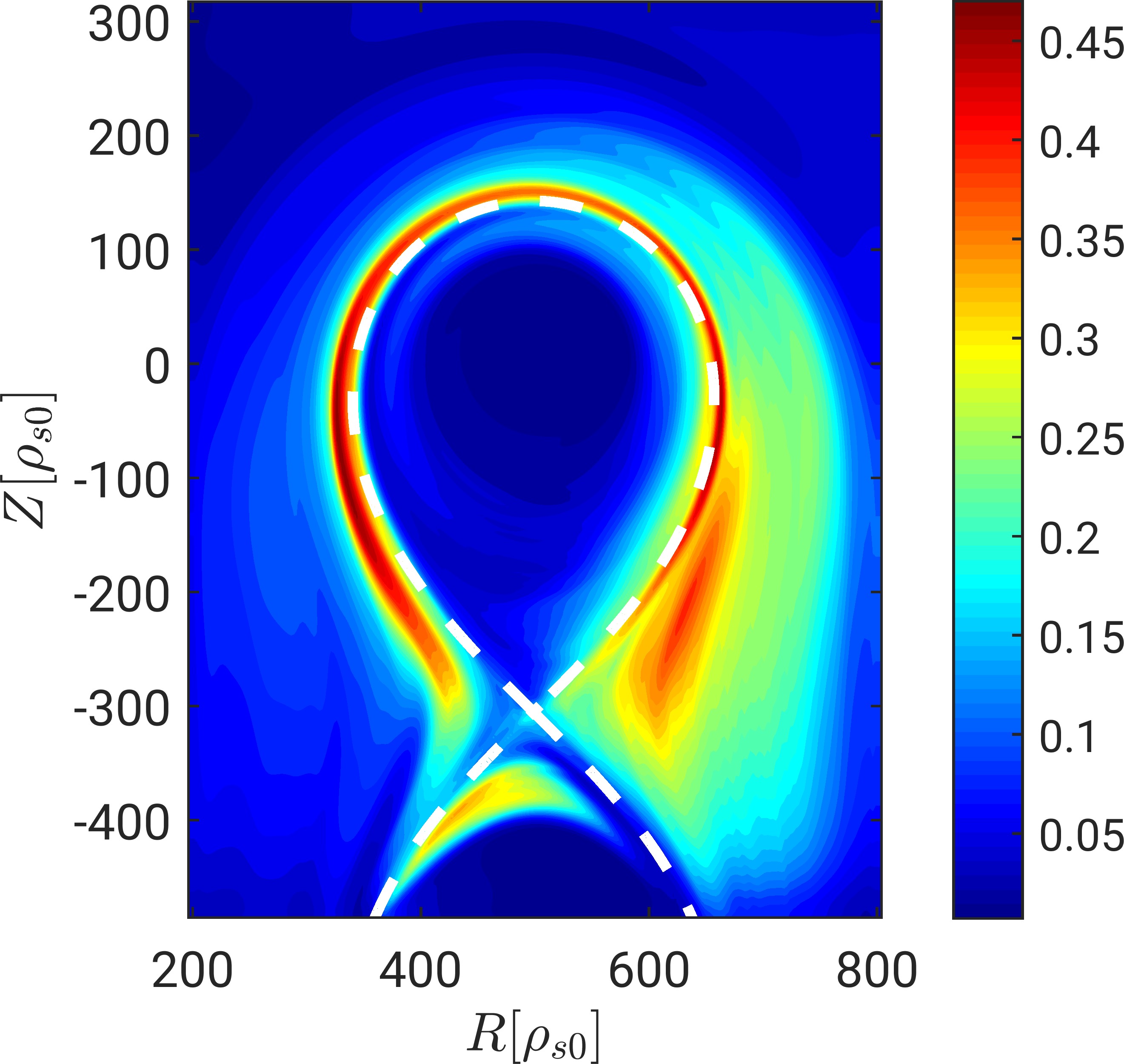}}
    \caption{Equilibrium density (a), equilibrium electrostatic potential (b), a snapshot of the relative density fluctuations (c) and the normalized standard deviation of the density fluctuations (d), for the simulation with $s_{T0} = 0.15$ and $\nu_0 = 0.2$. The dashed white line represents the separatrix.}
    \label{fig:snapshots}
\end{figure}

An example of typical simulation results is shown in Fig.~\ref{fig:snapshots} (more precisely, we consider the case $s_{T0} = 0.15$ and $\nu_0 = 0.2$). We note that the equilibrium density $\bar{n}$ is approximately a factor of 20 larger in the core than in the near SOL and a factor of 100 larger than in the far SOL (for any quantity $f$, we define its equilibrium value $\bar{f}$ as its time and toroidal average and the fluctuating component as $\tilde{f}=f-\bar{f}$). 
The equilibrium electrostatic potential $\bar{\phi}$ is positive in the SOL, while it drops and becomes negative inside the LCFS. 
The relative fluctuations of the density $\tilde{n}/\bar{n}$, shown by a typical snapshot, reveal that turbulence develops at the tokamak edge and propagates to the near SOL, with a strong interplay between these two regions.
In agreement with experimental observations \citep{terry2003,garcia2007,tanaka2009,ippolito2011}, the low-field side (LFS) of the far SOL is characterized by the presence of blobs, coherent radially propagating structures, whose dynamics in GBS simulations is analysed by~\citet{nespoli2017,paruta2019,beadle2020}. 
Indeed, as revealed by the study of standard deviation of the density fluctuations, the SOL is characterized by large fluctuations with amplitude comparable to the equilibrium quantities, as in the experiments \citep{horacek2005,boedo2009,kube2018}, while the level of density fluctuations in the core is very low, approximately 1~\%, also in agreement with experimental observations~\citep{fontana2017}.

By varying the heat source and the collisionality through the parameters $s_{T0}$ and $\nu_0$, respectively, three different turbulent regimes are identified in our simulations: (i) a regime of developed turbulent transport, which we link to the low-confinement mode (L-mode) of tokamak operation, discussed in \S\ref{sec:interchange}, (ii) a regime of suppressed turbulent transport, with similarities to the high-confinement mode (H-mode), discussed in \S\ref{sec:kh}, and (iii) a regime of degraded confinement with catastrophically high turbulent transport, which we associate to the crossing of the density limit and discuss in \S\ref{sec:density_limit}. While the transition from the developed to the suppressed transport regime is rather sharp, the transition to the degraded confinement regime is gradual.

\begin{figure}
    \centering
    \subfloat[]{\includegraphics[width=0.3\textwidth]{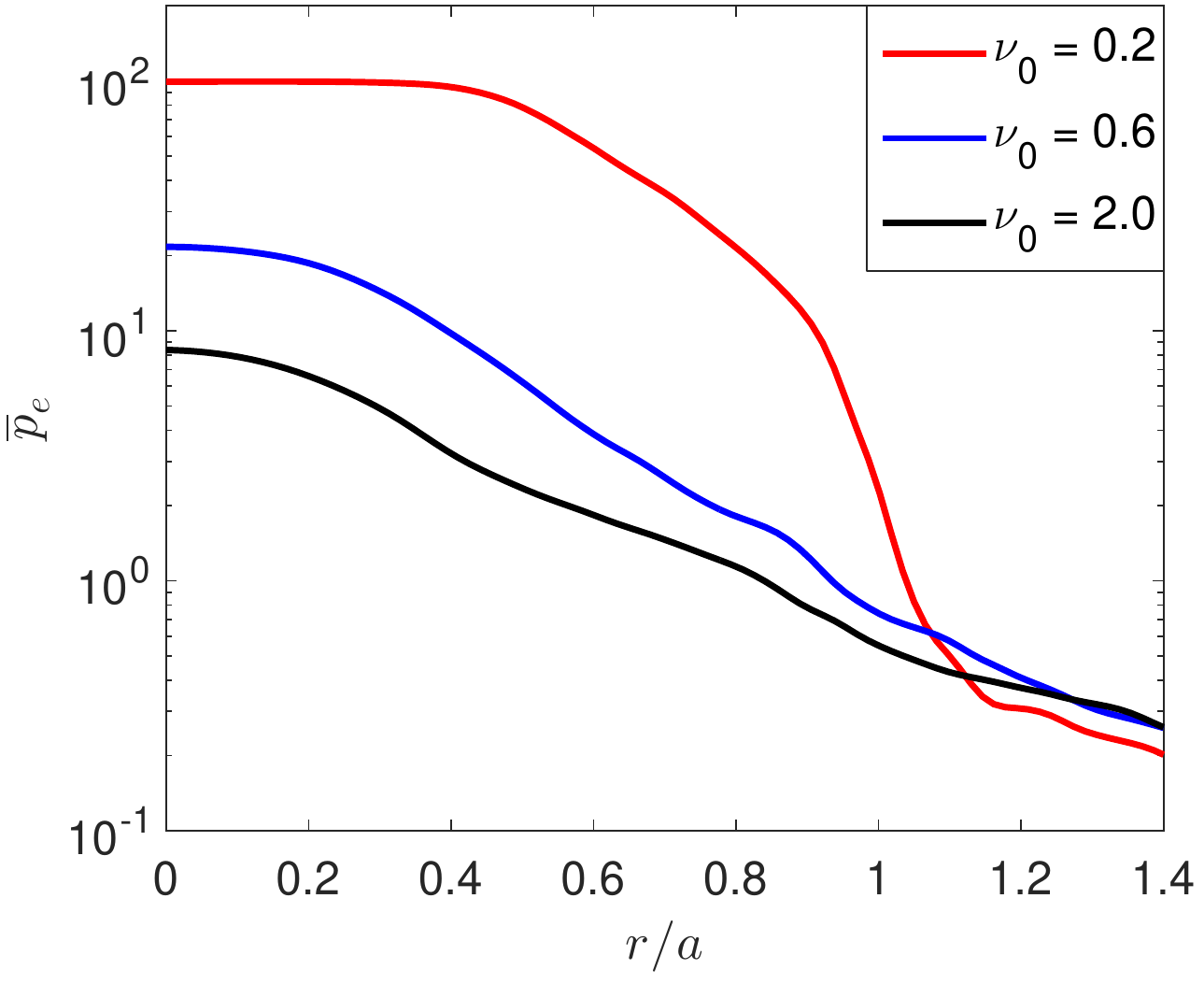}}\quad
    \subfloat[]{\includegraphics[width=0.3\textwidth]{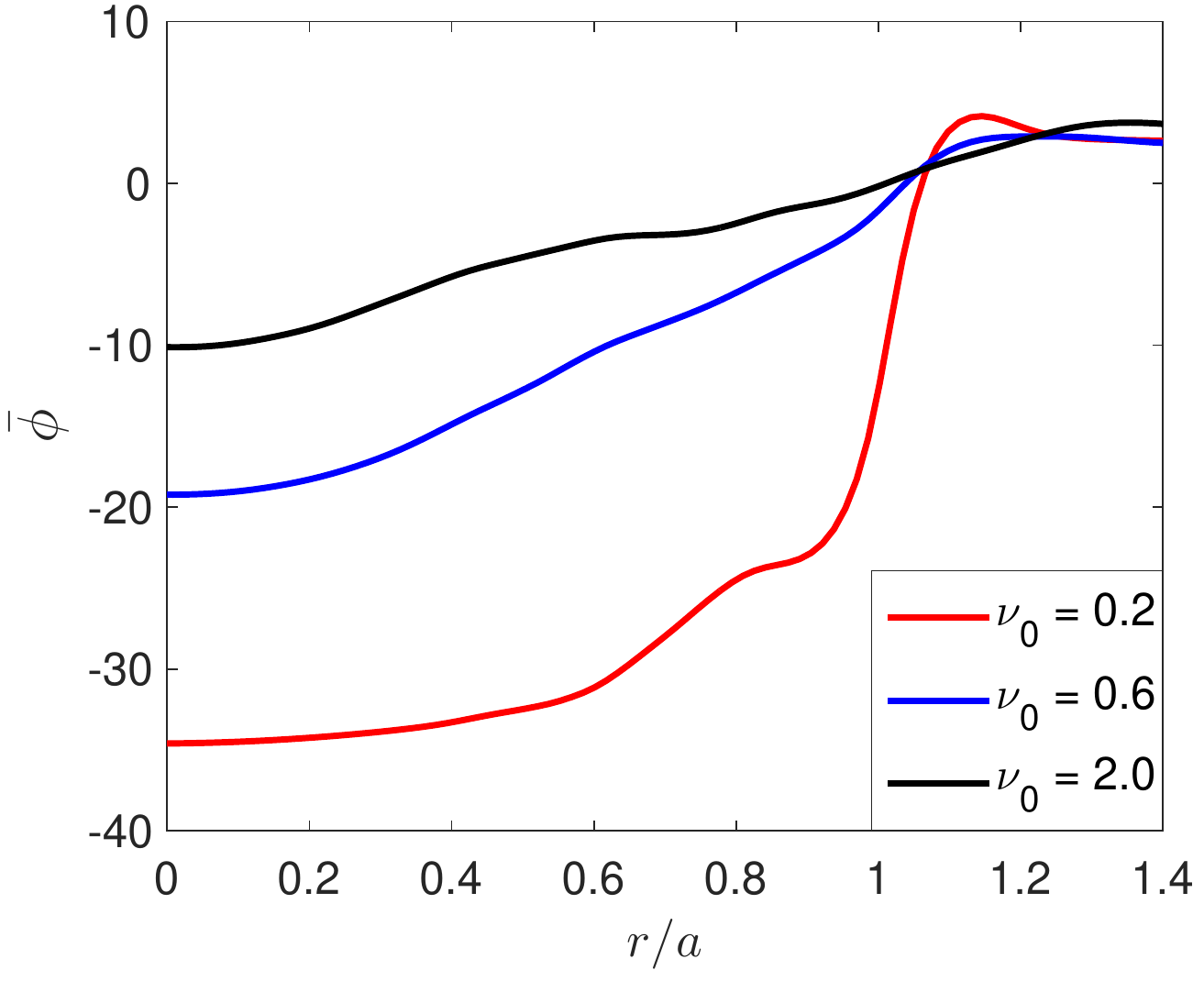}}\quad
    \subfloat[]{\includegraphics[width=0.3\textwidth]{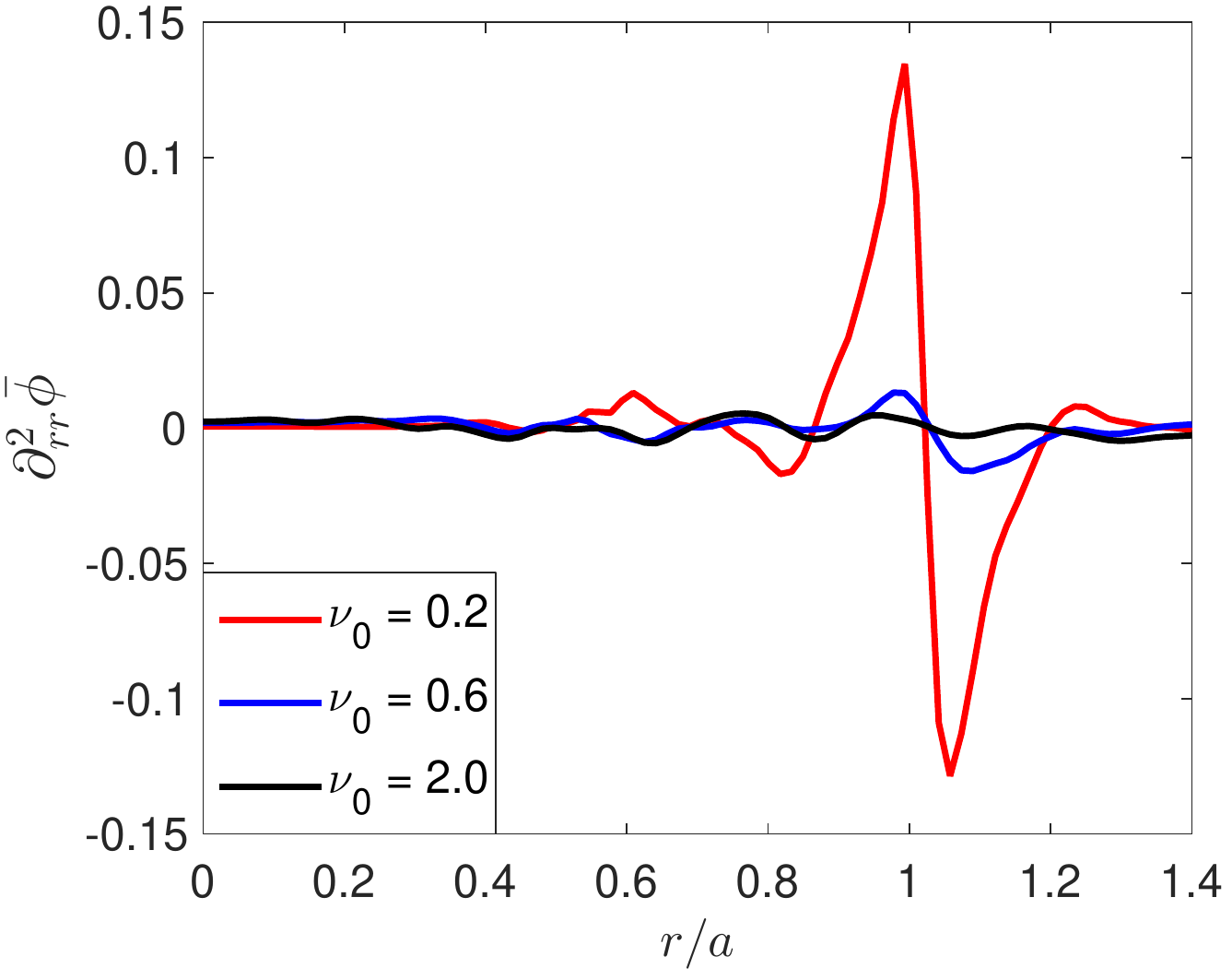}}
    \caption{Radial profiles at the LFS midplane of the equilibrium pressure (a), equilibrium electrostatic potential (b), and equilibrium $\mathbf{E}\times\mathbf{B}$ shear (c) for the simulation with $s_{T0}=0.15$ and $\nu_0=0.2$ (suppressed transport regime), $\nu_0=0.6$ (developed transport regime), and $\nu_0=2.0$ (degraded confinement regime). The radial coordinate is normalized to the radial position $a$ of the separatrix at the midplane.}
    \label{fig:profiles}
\end{figure}

Typical radial profiles at the LFS midplane of the equilibrium pressure, electrostatic potential and $\mathbf{E}\times\mathbf{B}$ shear are shown in Fig.~\ref{fig:profiles} for the three regimes. We consider the simulations with $s_{T0}=0.9$ and $\nu_0=0.2$ (suppressed transport regime), $\nu_0=0.6$ (developed transport regime), and $\nu_0=2.0$ (degraded confinement regime). In the suppressed transport regime, the electrostatic potential drops significantly inside the separatrix, generating a strong $\mathbf{E}\times\mathbf{B}$ shear across it.
This is associated to a steep gradient in the density, electron and ion temperatures.
With respect to the suppressed transport regime, in the developed transport regime  the electrostatic potential across the separatrix is flatter, the equilibrium $\mathbf{E}\times\mathbf{B}$ shear is reduced, transport due to turbulence is larger and, consequently, the density and temperature gradient at the tokamak edge is significantly lower. In the degraded confinement regime, turbulent transport is extremely large, leading to a flat profile of density, temperature and electrostatic potential. We note that analogous transitions can be observed by varying the heat source while keeping $\nu_0$ constant.

\begin{figure}
    \centering
    \subfloat[Suppressed transport]{\includegraphics[height=0.22\textheight]{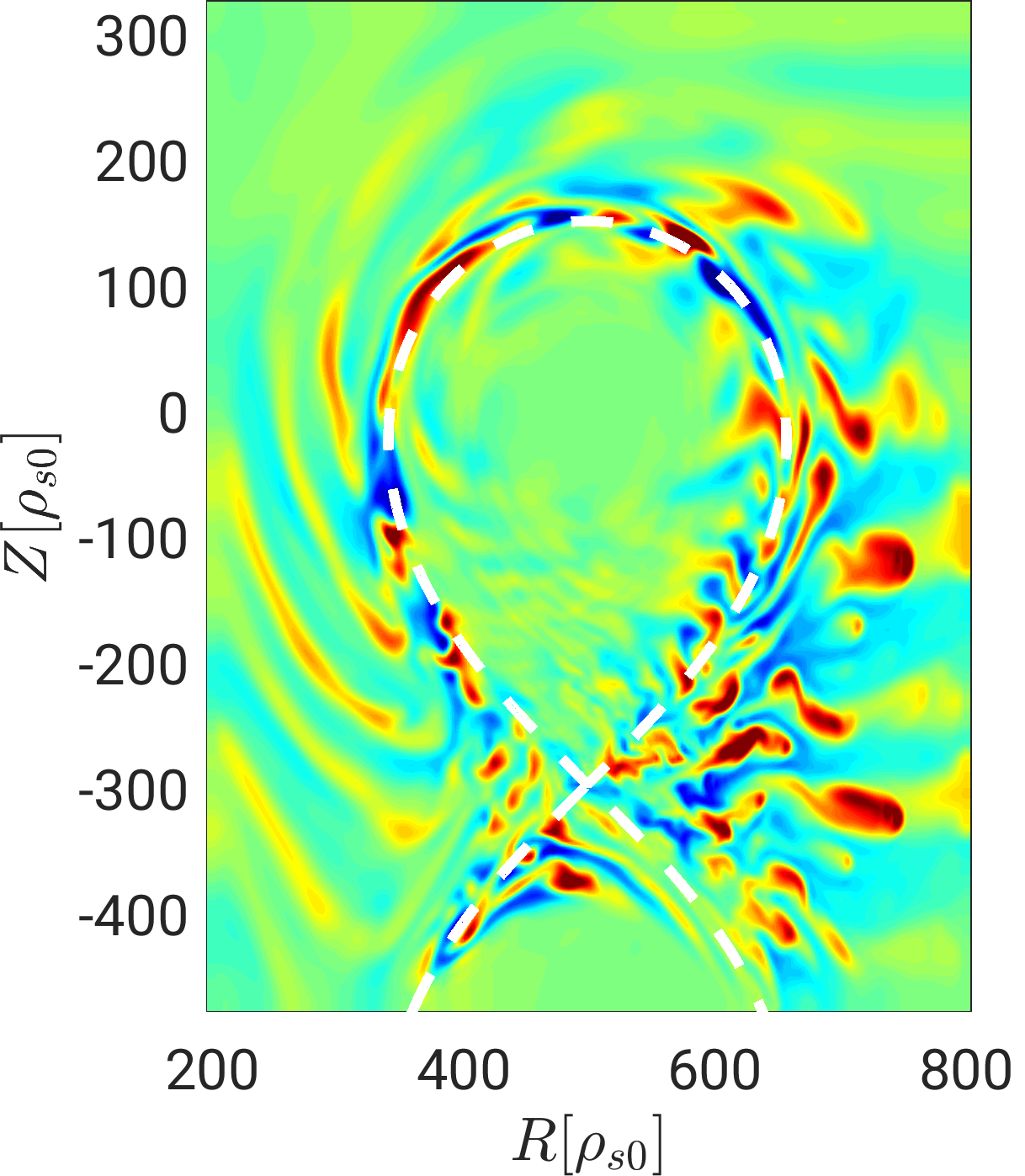}}\ 
    \subfloat[Developed transport]{\includegraphics[height=0.22\textheight]{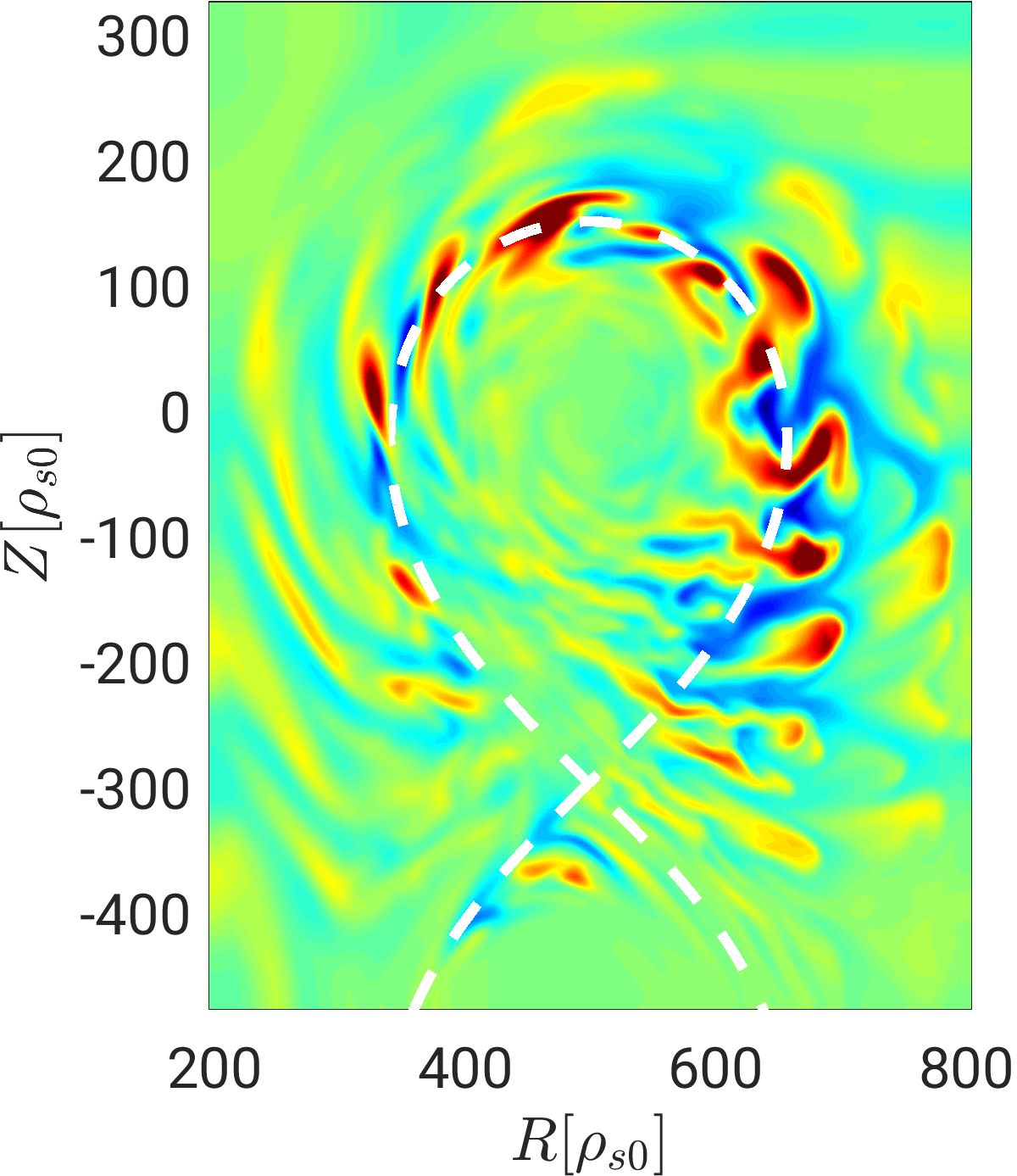}}\ 
    \subfloat[Degraded confinement]{\includegraphics[height=0.22\textheight]{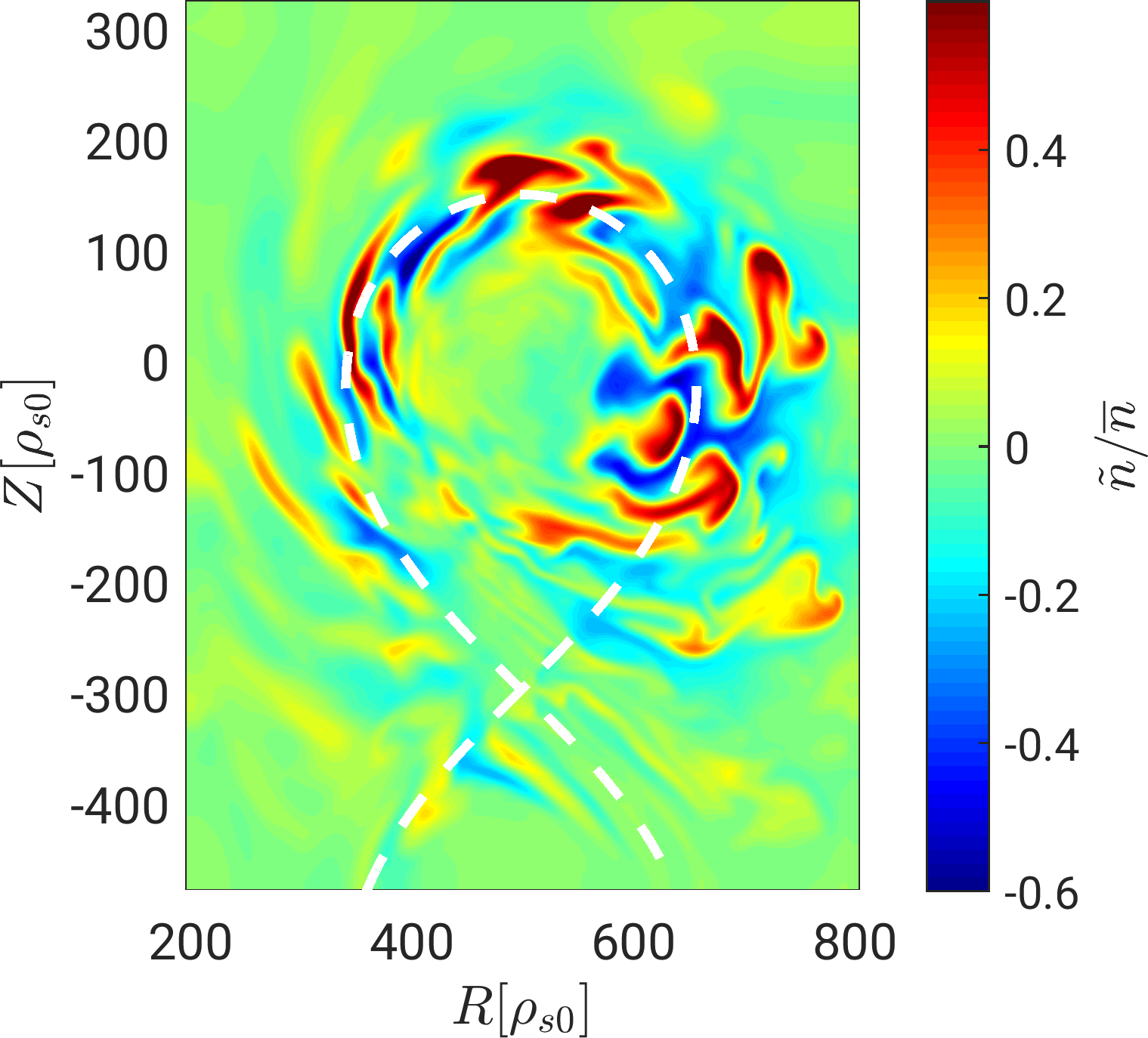}}\\
    \subfloat[Suppressed transport]{\includegraphics[height=0.22\textheight]{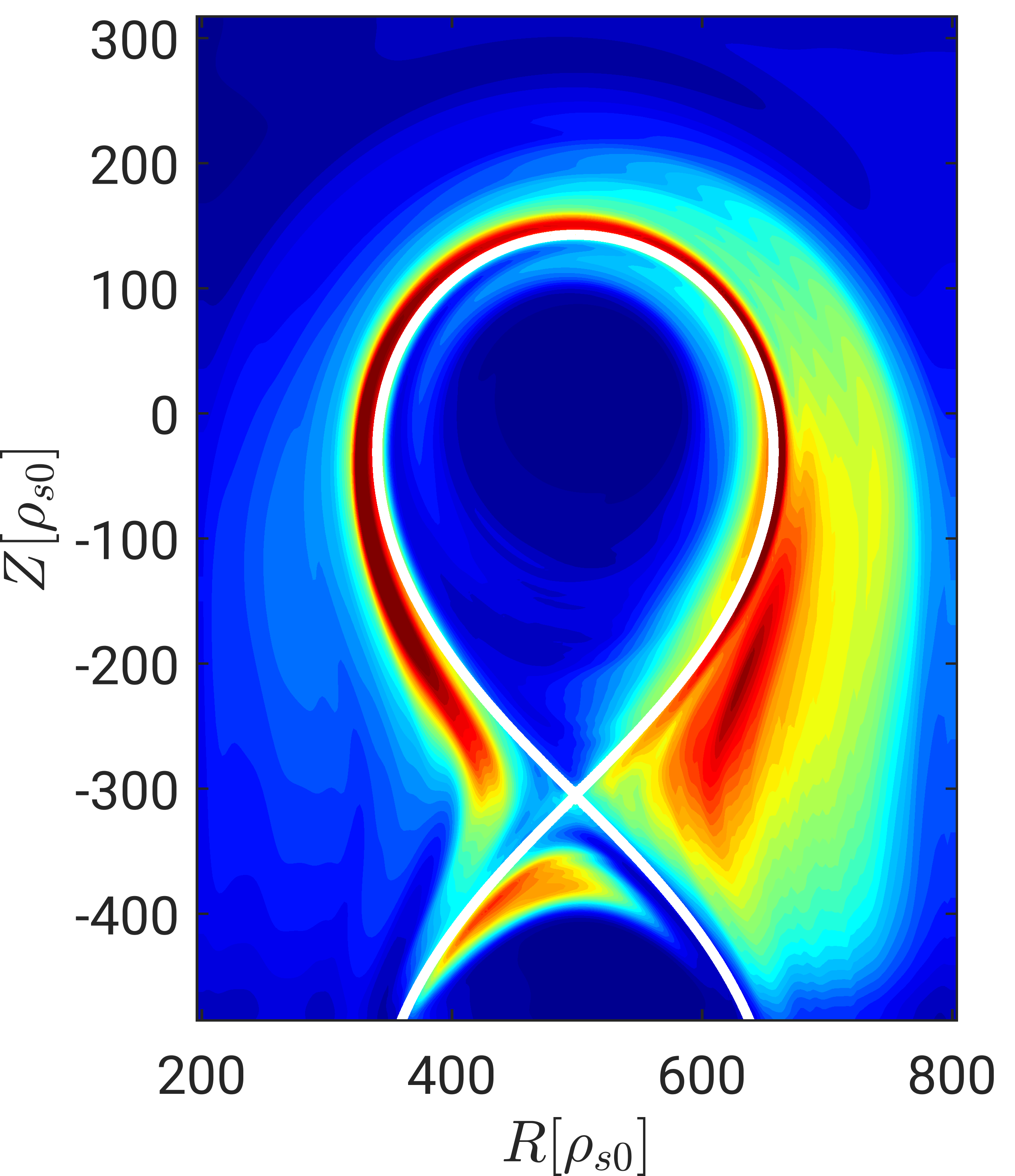}}\ 
    \subfloat[Developed transport]{\includegraphics[height=0.22\textheight]{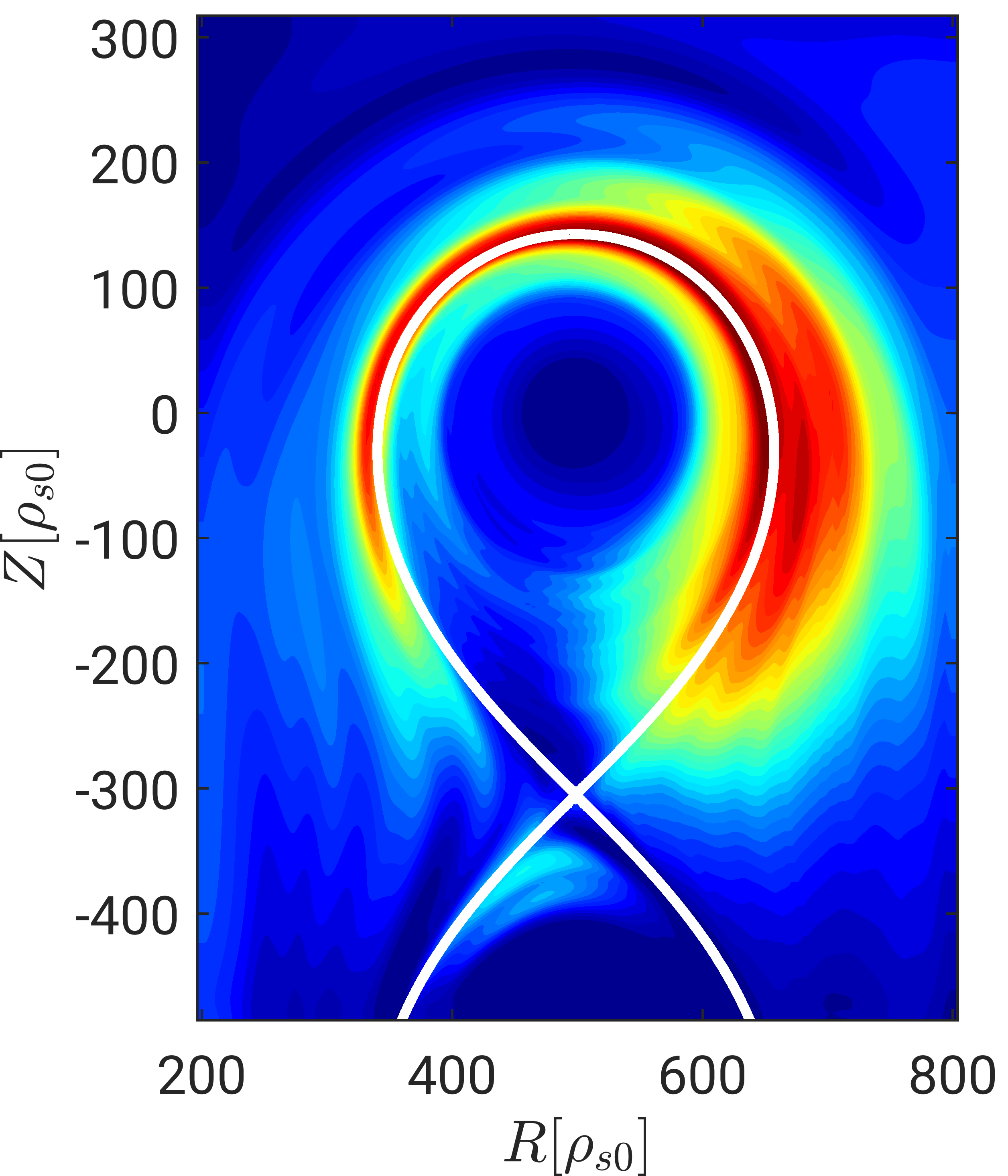}}\ 
    \subfloat[Degraded confinement]{\includegraphics[height=0.22\textheight]{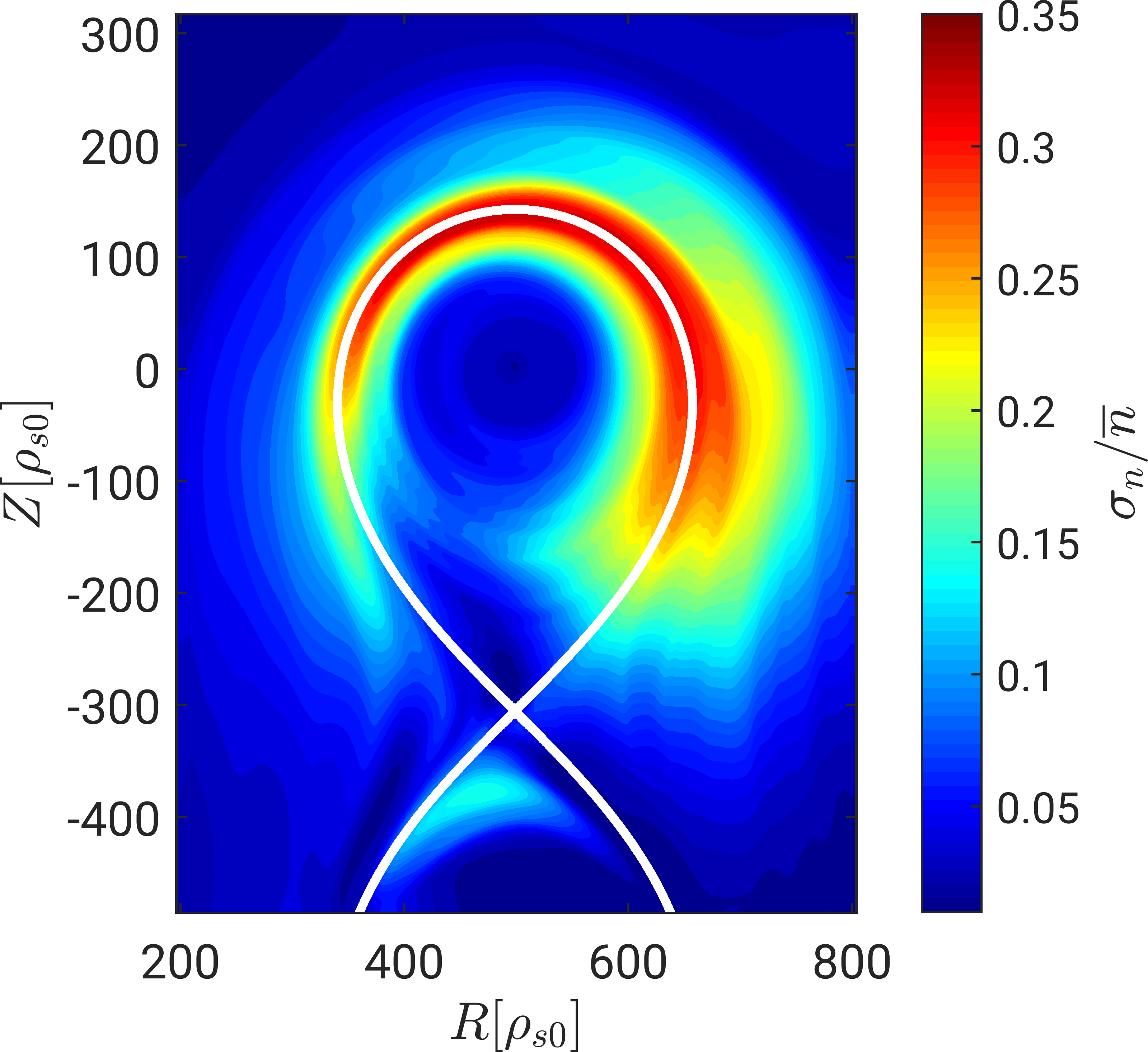}}
    \caption{Typical snapshot of the relative density fluctuations (top raw) and normalized standard deviation of the density fluctuations (bottom raw) for three simulations with $s_{T0}=0.15$ in the suppressed transport regime, $\nu_0=0.2$  [(a) and (d)], developed transport regime, $\nu_0=0.6$ [(b) and (e)], and degraded confinement regime, $\nu_0=2.0$ [(c) and (f)].}
    \label{fig:den_limit}
\end{figure}

Typical snapshots of plasma turbulence in the three transport regimes can be seen in Fig.~\ref{fig:den_limit}, where the relative density fluctuations and the corresponding normalized standard deviation are shown for the three simulations we are considering.
In the case of $\nu_0=0.2$, turbulence is localized near the separatrix and, as a consequence of being sheared apart by the strongly varying $\mathbf{E}\times\mathbf{B}$ radial profile, turbulent structures are elongated along the $\nabla\chi$ direction, effectively reducing the cross-field transport.  
The radial extension of turbulent structures is larger for $\nu_0=0.6$ and $\nu_0=2.0$. In particular, at $\nu_0=2.0$, turbulent structures penetrate into the core region. This is in agreement with experimental observations of density fluctuations when the density limit is approached \citep{labombard2001}. 
In addition, in the case of $\nu_0=0.2$, density fluctuations are generated both at the LSF and high-field side (HFS), while, in the other two cases, turbulence mainly develops at the LFS.

\begin{figure}
    \centering
    \includegraphics[scale=0.65]{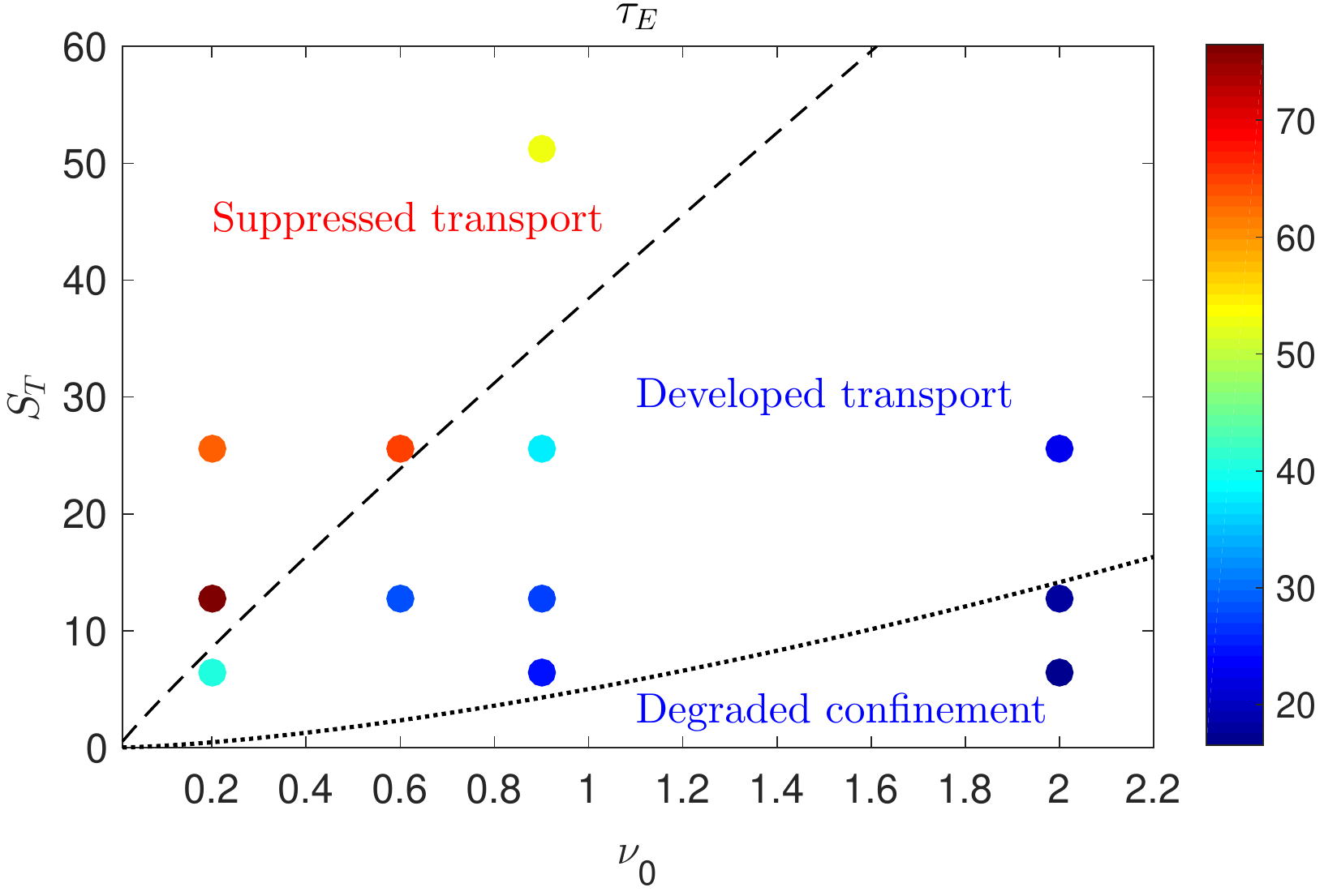}
    \caption{Electron energy confinement time for the set of simulations performed for the present study. The dashed black line represents the heat source threshold to access the suppressed transport regime (derived in \S\ref{sec:power_threshold}, see Eq.~\eqref{eqn:temp_source}), while the dotted black line represents the heat source threshold to access the degraded confinement regime (derived in \S\ref{sec:density_threshold}, see Eq.~\eqref{eqn:temperature_den_limit}).}
    \label{fig:conf_time}
\end{figure}

In order to highlight the difference on the confinement properties between the different regimes, we compute the electron energy confinement time, $\tau_E=\frac{3}{2}\int_{A_\text{LCFS}} \bar{p}_e\,\mathrm{d}R\mathrm{d}Z/\int_{A_\text{LCFS}} s_p\,\mathrm{d}R\mathrm{d}Z$, for the set of simulations considered in the present study, at different values of $S_{T}$ and $\nu_0$ (see Fig.~\ref{fig:conf_time}). 
At a given $\nu_0$ or $S_T$, we note that the simulations in the suppressed transport regime have a higher energy confinement time than the simulations in the developed transport regime. 
For this reason, we also refer to the developed transport regime as the L-mode and to the suppressed transport regime as the H-mode. 
The energy confinement time increases by a factor of two from the L-mode to the H-mode, as observed in the experiments.
In addition, as a consequence of the larger fluctuations, the energy confinement time is lower in the degraded confinement regime than in the developed transport regime.

The detailed analysis of the three regimes is reported in~\S\ref{sec:transport_regimes}. The power threshold to access the suppressed transport regime and the degraded confinement regime, both displayed in Fig.~\ref{fig:conf_time} as a function of $\nu_0$, are discussed in~\S\ref{sec:transitions}. 

\section{Turbulent transport regimes at the tokamak edge\label{sec:transport_regimes}}

In this section, we analyse separately the three transport regimes revealed by our simulations. The mechanisms driving turbulence are studied and an analytical expression of the edge equilibrium pressure gradient length is derived for the different transport regimes.    

\subsection{Developed transport regime (L-mode)}
\label{sec:interchange}

\begin{figure}
    \centering
    \subfloat[]{\includegraphics[height=0.23\textheight]{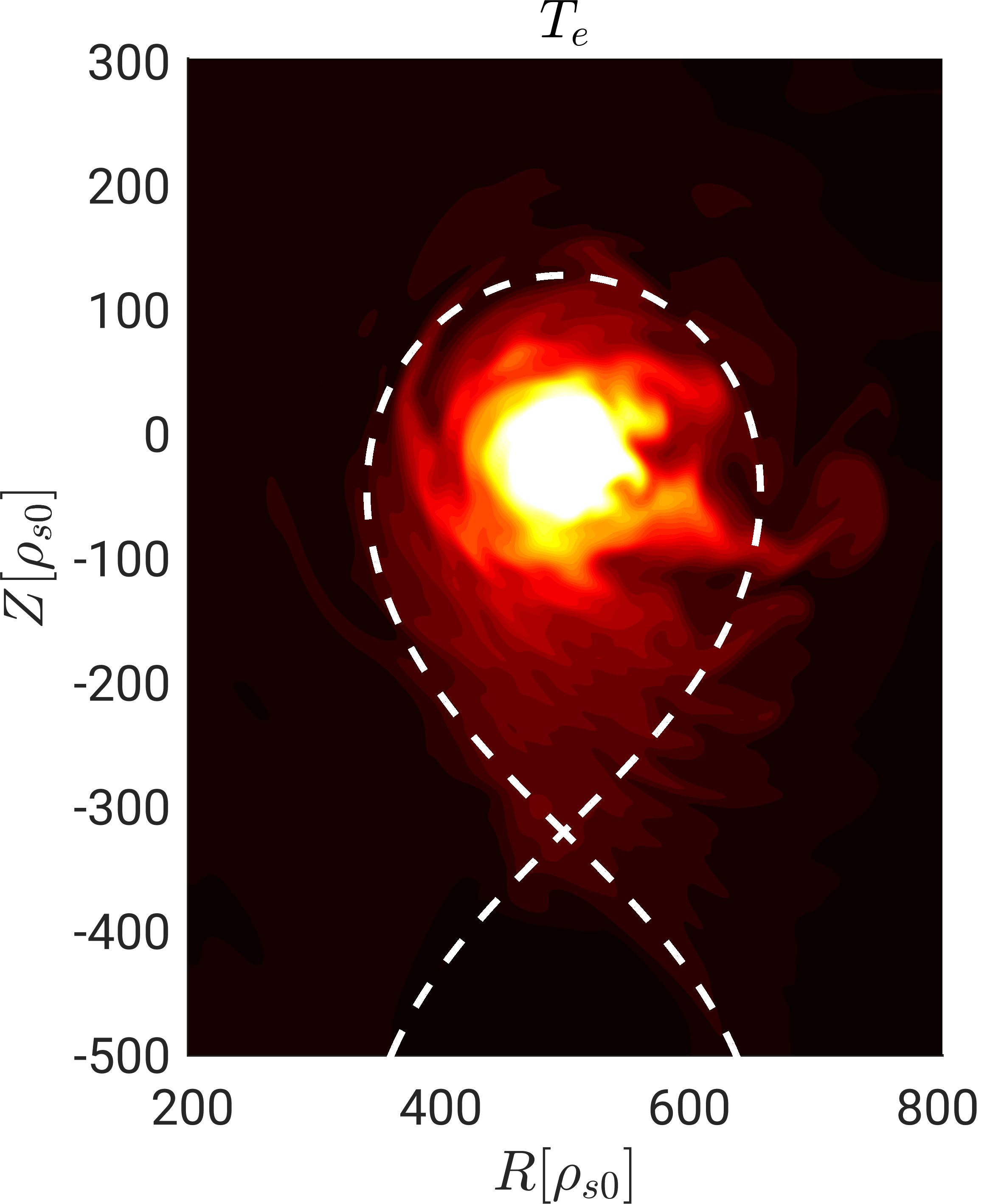}}\,
    \subfloat[]{\includegraphics[height=0.23\textheight]{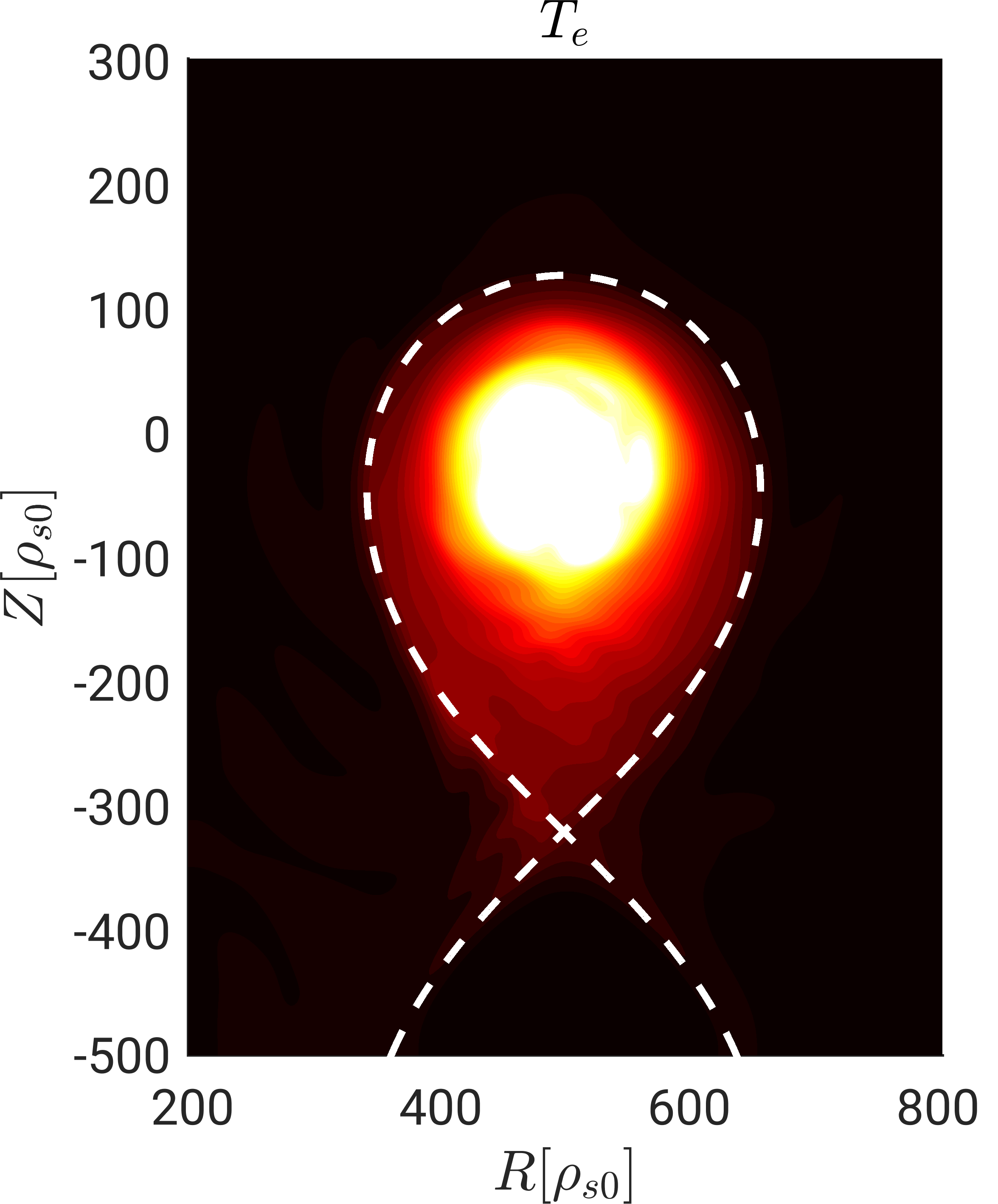}}\,
    \subfloat[]{\includegraphics[height=0.23\textheight]{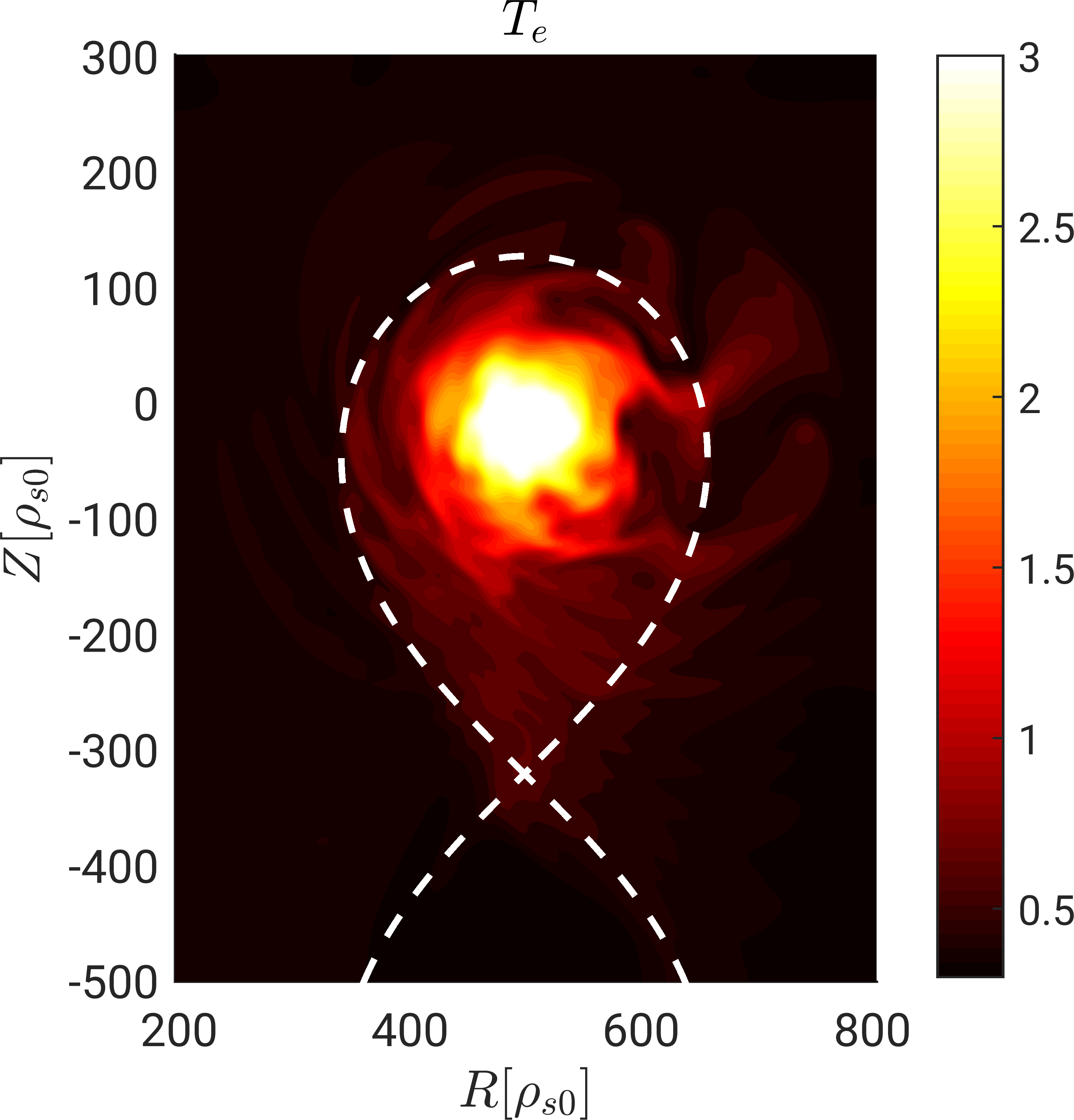}}
    \caption{A typical snapshot of the electron temperature for the simulation with $s_{T0}=0.075$ and $\nu_0=0.9$ (a). Snapshots of simulations with the same parameters but the interchange drive $C(p_e+\tau p_i)$ term in Eq.~\eqref{eqn:vorticity} toroidally averaged (b), and with the KH drive term $\rho_*^{-1} \bigl[\phi,\omega\bigr]/B$ in Eq.~\eqref{eqn:vorticity} toroidally averaged (c).}
    \label{fig:no_int}
\end{figure}

We start by considering the regime of developed transport, which we associate to the L-mode. In this regime, shown by our simulations at intermediate heat source values and intermediate values of collisionality (see Fig.~\ref{fig:conf_time}), the shear flows is negligible and turbulent transport results from the nonlinear development of interchange-driven electrostatic ballooning modes \citep{Mosetto2013}. This can be verified by removing the interchange drive from the simulations, i.e. by toroidally averaging the term proportional to $C(p_e+\tau p_i)$ in Eq.~\eqref{eqn:vorticity}. 
The result of this test is displayed in Fig.~\ref{fig:no_int}, where a snapshot of the electron temperature, with and without the interchange drive, is shown for a simulation in the L-mode regime ($s_{T0} = 0.075$ and $\nu_0 = 0.9$). Plasma turbulence is strongly suppressed when the term $C(p_e+\tau p_i)$ is toroidally averaged and, as a consequence, an increase of the equilibrium temperature and pressure gradients is observed. On the other hand, turbulent structures and plasma profiles do not change significantly when the Reynolds stress, i.e. the term $\rho_*^{-1}\bigl[\phi,\omega\bigr]/B$ appearing in Eq.~\eqref{eqn:vorticity}, is toroidally averaged (see Fig.~\ref{fig:no_int}). This shows that the $\mathbf{E}\times\mathbf{B}$ shear and the KH instability do not play a major role in the developed transport regime.  

In order to provide an analytical estimate of the pressure gradient length in the edge, we follow a procedure similar to the one described by \citet{Ricci2008}, and we balance the perpendicular heat flux crossing the separatrix with the heat source integrated over the volume inside the LCFS.
The simulations show that the equilibrium cross-field heat flux near the separatrix is negligible with respect to the turbulent one, $\bar{p}_e\partial_\chi\bar{\phi} \ll \overline{\tilde{p}_e\partial_\chi\tilde{\phi}}$ ($\partial_\chi$ denotes the derivative along $\nabla\chi$). 
Therefore, we focus on the perpendicular turbulent transport, ${q_\psi \simeq \overline{\tilde{p}_e\partial_\chi\tilde{\phi}}}$, at the LCFS.
The quantity $\partial_\chi\tilde{\phi}$ is estimated from the leading terms of the linearized electron pressure equation, which is obtained by linearizing and summing Eqs.~\eqref{eqn:density}~and~\eqref{eqn:electron_temperature},  
\begin{equation}
    \label{eqn:lin_pressure}
    \partial_t \tilde{p}_e \sim \rho_*^{-1} \partial_\psi \bar{p}_e \partial_\chi\tilde{\phi}\,,
\end{equation}
(the curvature and parallel gradient terms appearing in Eqs.~\eqref{eqn:density}~and~\eqref{eqn:electron_temperature} are significantly smaller than the terms we retain). 
In Eq.~\eqref{eqn:lin_pressure}, we estimate the time derivative as the growth rate of the ballooning instability driving the transport, $\gamma_i=\sqrt{2 \bar{T}_e/(\rho_*L_p)}$. We also approximate $\partial_\psi \bar{p}_e \simeq \bar{p}_e/L_p$ being $L_p$ the equilibrium pressure gradient length. 
The resulting expression of $\partial_\chi\tilde{\phi}$ can then be used to evaluate the cross-field interchange-induced heat flux as
\begin{equation}
    \label{eqn:transport_intermediate}
    q_{\psi,i}\sim \rho_*\gamma_i\frac{\tilde{p}_e^2}{\bar{p}_e}L_p\,.
\end{equation}
The amplitude of the pressure fluctuations appearing in Eq.~\eqref{eqn:transport_intermediate} can be estimated by observing that the growth of the linearly unstable modes saturates when the instability drive is removed from the system, i.e. $k_\psi \tilde{p}_e \sim \bar{p}_e/L_p$ \citep{Ricci2013,Ricci2008}.  
The perpendicular heat flux is then written as
\begin{equation}
    \label{eqn:transport}
    q_{\psi,i}\sim\rho_*\frac{\gamma_i}{k_{\psi,i}^2}\frac{\bar{p}_e}{L_p}\,.
\end{equation}
Non-local linear calculations show that $k_{\psi,i}\simeq\sqrt{k_{\chi,i}/L_p}$ \citep{Ricci2008}.
The poloidal wavenumber of the ballooning instability $k_{\chi,i}$ can then be obtained by balancing the interchange driving and the parallel current terms in Eq.~\eqref{eqn:vorticity}. In the parameter regime of our simulations, turbulence is driven by resistive ballooning mode \citep{Mosetto2013}. In this case, the resistivity limits the parallel current \citep{Halpern2013a,Halpern2014}. This leads to  $k_{\chi,i} = (\bar{n}\nu q_{95}^2\sqrt{\gamma_i})^{-1/2}$. As a consequence, Eq.~\eqref{eqn:transport} becomes
\begin{equation}
    \label{eqn:transport_int}
    q_{\psi,i}\sim \rho_*^{1/4}\sqrt{\bar{n}\nu q_{95}^2}\biggl(\frac{2\bar{T}_e}{L_p}\biggr)^{3/4}\bar{p}_e\,.
\end{equation}

In order to derive the pressure scale length, we note that the heat source integrated over the poloidal plane inside the LCFS corresponds, approximately, to the perpendicular turbulent heat flux crossing the LCFS on a poloidal plane,
\begin{equation}
    \label{eqn:flux_balance}
    S_p(R,Z) \simeq \oint_{\text{LCFS}} q_{\psi,i}(R,Z)\,\mathrm{d}l \,.
\end{equation}
In order to perform the integral on the right-hand side of Eq.~\eqref{eqn:flux_balance}, we note that turbulent transport is driven by ballooning modes that develop in the bad-curvature region (see Fig.~\ref{fig:no_int}). As a consequence, we assume that $q_{\psi,i}(R,Z)$ has a constant value at the LCFS on the LFS and vanishes at the HFS, i.e.
\begin{equation}
    \label{eqn:balance_approx}
    S_p\sim\frac{L_\chi}{2}q_{\psi,i}\,,
\end{equation}
where $L_\chi= \oint_{\text{LCFS}} \mathrm{d}l$ is the length of the LCFS poloidal circumference.
The edge equilibrium pressure gradient length is derived by using Eqs.~\eqref{eqn:transport_int}~and~\eqref{eqn:balance_approx}, that is
\begin{equation}
    \label{eqn:lp_int}
    L_{p,i} \sim \biggl[\frac{\rho_*}{2}(\nu q_{95}^2 \bar{n})^2\biggl(\frac{L_\chi}{S_p}\bar{p}_e\biggr)^4\biggr]^{1/3}\bar{T}_e\,, 
\end{equation}
where $\bar{n}$, $\bar{T}_e$, and $\bar{p}_e$ are evaluated at the LCFS.

\begin{figure}
    \centering
    \includegraphics[scale=0.65]{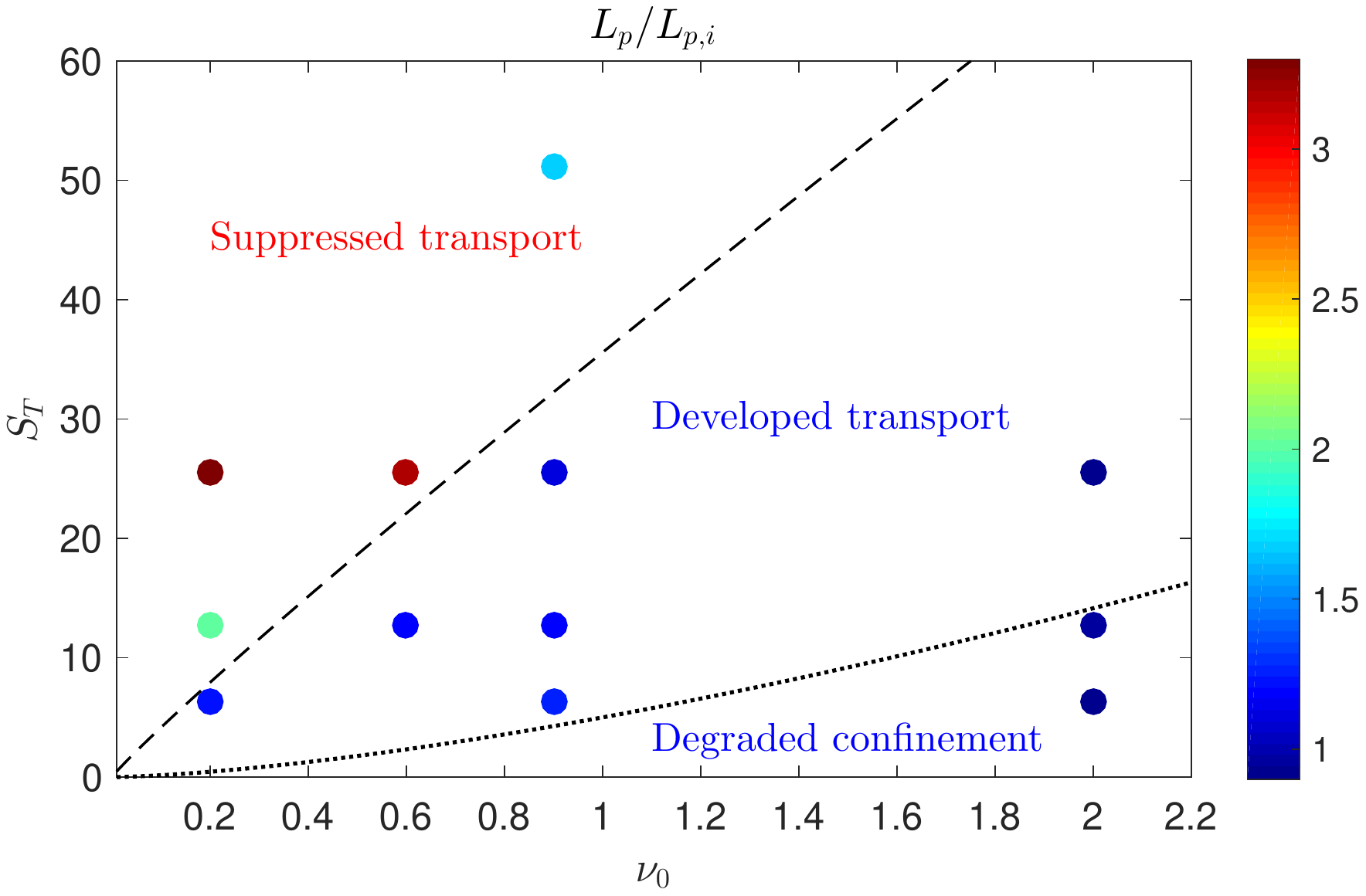}
    \caption{Ratio of $L_p$, the equilibrium pressure gradient length directly obtained from the simulations, to $L_{p,i}$, the estimate in Eq.~\eqref{eqn:lp_int} based on the assumption that transport is driven by the interchange instability, for all the simulations considered in the present study. The dashed black line represents the heat source threshold to access the suppressed transport regime, Eq.~\eqref{eqn:temp_source}, while the dotted black line the threshold to access the degraded confinement regime, Eq.~\eqref{eqn:temperature_den_limit}. }
    \label{fig:lp_int}
\end{figure}

In Fig.~\ref{fig:lp_int}, the ratio of $L_p$, the equilibrium pressure gradient length directly obtained from the simulations, to $L_{p,i}$, the interchange estimate in Eq.~\eqref{eqn:lp_int}, is displayed for the different values of $S_{T}$ and $\nu_0$ considered in the present study. At low values of $S_T$ and high $\nu_0$,  $L_p/L_{p,i}\simeq 1$, revealing a good agreement between the analytical estimate in Eq.~\eqref{eqn:lp_int} and the simulation result. Hence, in the developed transport regime (as well as in the degraded confinement regime, as discussed in \S\ref{sec:density_limit}), turbulence is driven by the interchange instability, being the effect of the shear flow not significant.
In addition, since the formation of a pedestal is not observed in this regime, we associate this parameter region to the L-mode. 
On the other hand, for high values of the heat source and low $\nu_0$, the pressure gradient length of the simulations is larger than the value predicted by Eq.~\eqref{eqn:lp_int}, $L_p/L_{p,i} > 1$, meaning that a mechanism different than the interchange instability is responsible for driving turbulent transport and setting the equilibrium pressure gradient length. 

It should be noted that, despite being still described as the result of the development of the interchange instability, transport might become catastrophically large at high values of $\nu_0$ and low heat source values, with turbulent eddies that extend from the edge towards the tokamak core, a behaviour that we associate to the crossing of the density limit and we identify as the degraded confinement regime.

\subsection{Suppressed transport regime (H-mode)}\label{sec:kh}

As shown in Fig.~\ref{fig:profiles}, the equilibrium edge electrostatic potential profile in the suppressed transport regime is significantly different from the one in the developed transport regime.  
This has strong consequences on the nature of turbulent transport. 
Hence, we focus on the mechanisms that set the $\bar{\phi}$ profile.

Inside the LCFS, the radial electric field is proportional to the ion pressure gradient, $\partial_r \bar\phi \sim -\partial_r \bar{p}_i/\bar{n}$, as experimentally observed (see, e.g., \citet{schirmer2006} and \citet{mcdermott2009}) and theoretically explained (see, e.g., \citet{Zhu2017}). 
On the other hand, ambipolarity of the plasma flow at the sheath imposes that the electrostatic potential is proportional to the electron temperature, $\bar{\phi} \sim \Lambda \bar{T}_e$, in the SOL, as discussed by \citet{stangeby2000}~and~\citet{Loizu2013}. 
Therefore, $\bar{\phi}$ radially increases as one moves from the magnetic axis towards the LCFS ($-\partial_r \bar{p}_i/\bar{n} > 0$) and then decreases from the LCFS towards the far SOL ($\partial_r\bar{T}_e<0$). It follows that  $\bar{\phi}$ peaks near the separatrix (see Fig.~\ref{fig:profiles}~(b)). 
As $s_{T0}$ increases or $\nu_0$ decreases, both $T_e$ and $T_i$ increase and, as a consequence, also the $\mathbf{E}\times\mathbf{B}$ shear flow across the LCFS increases (see Fig.~\ref{fig:profiles}~(c)). 
Because of the $\mathbf{E}\times\mathbf{B}$ shear, the turbulent eddies in the edge resulting from the interchange instability are sheared along the $\nabla \chi$ direction. Furthermore, when the shearing rate, $\rho_*^{-1}\partial_{rr}^2\bar{\phi}$, is comparable to $\gamma_i$, ballooning turbulence is nonlinearly suppressed \citep{Burrell1997,Terry2000}. 
At the same time, the $\mathbf{E}\times\mathbf{B}$ shear provides the drive of the KH instability through the Reynolds stress \citep{Myra2016}. 
Indeed, our simulations show that, for values sufficiently high of the heat source, when $\rho_*^{-1}\partial_{rr}^2\bar{\phi} > \gamma_i$, the interchange instability is suppressed in the edge and the KH instability becomes the primary instability driving the turbulent transport \citep{Rogers2005,Myra2016}.

\begin{figure}
    \centering
    \subfloat[]{\includegraphics[height=0.23\textheight]{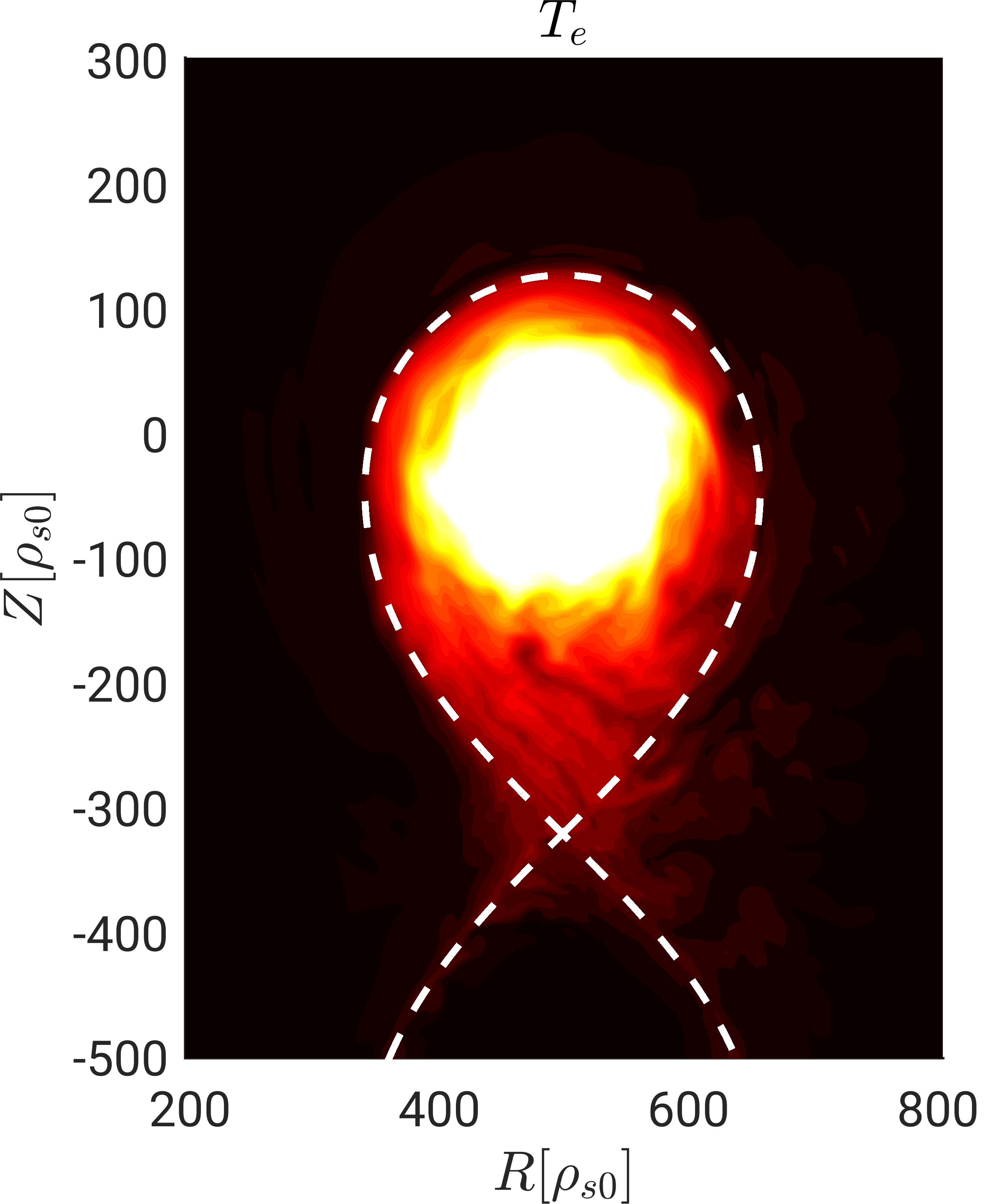}}\,
    \subfloat[]{\includegraphics[height=0.23\textheight]{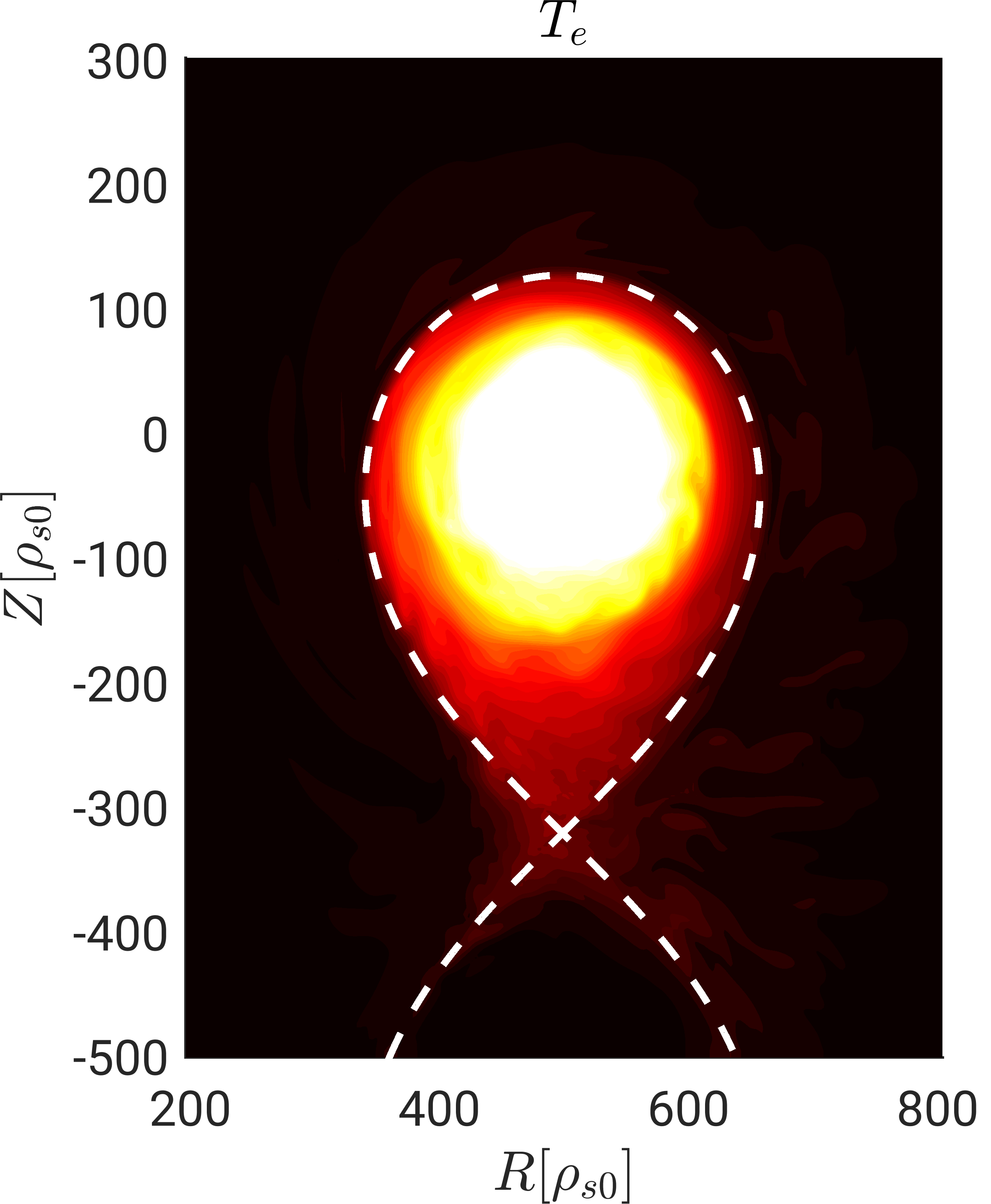}}\,
    \subfloat[]{\includegraphics[height=0.23\textheight]{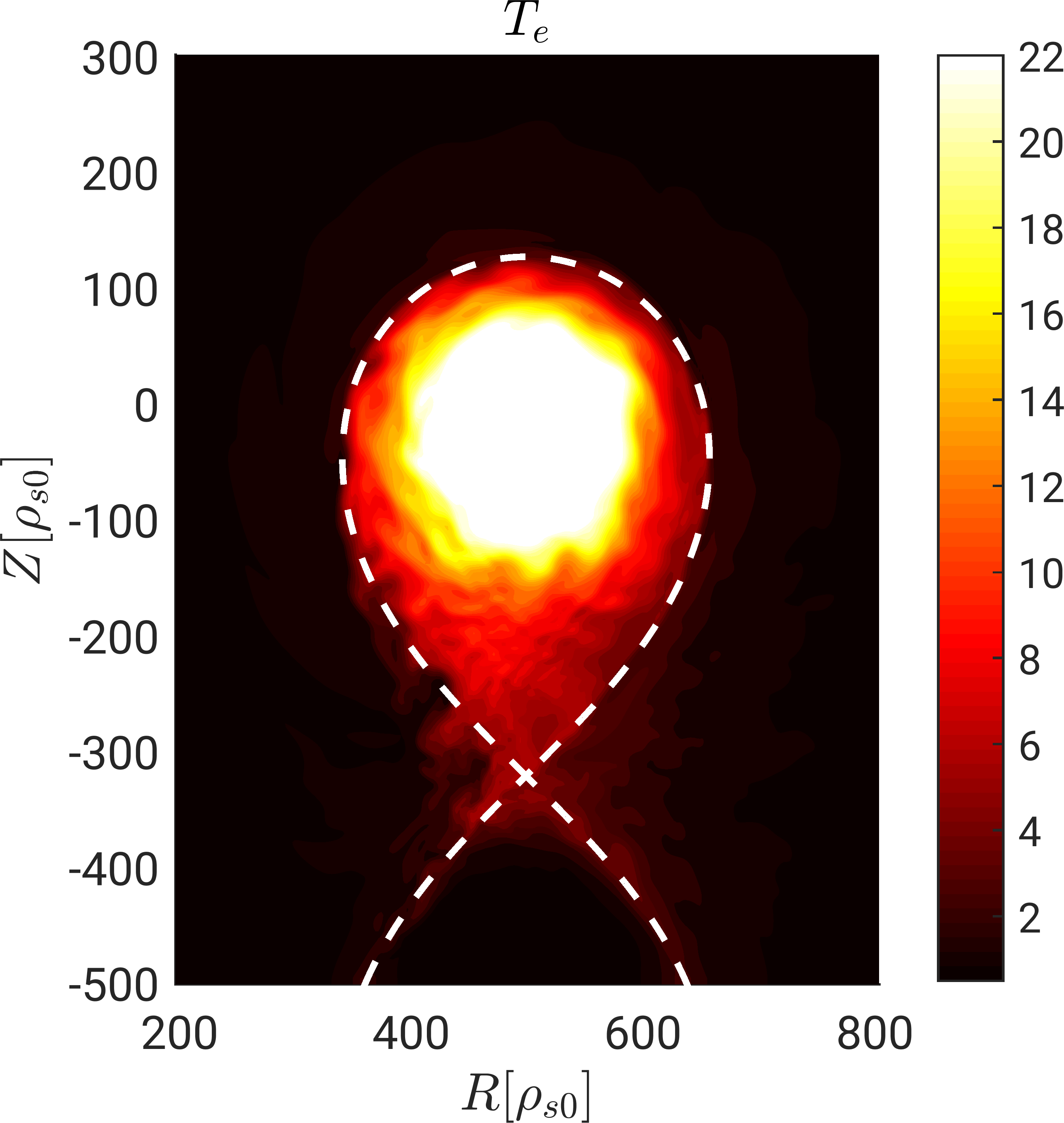}}
    \caption{A typical snapshot of the electron temperature for the simulation with $s_{T0}=0.6$ and $\nu_0=0.9$ (a). Snapshots of simulations with the same parameters but the KH drive term $\rho_*^{-1} \bigl[\phi,\omega\bigr]/B$  in Eq.~\eqref{eqn:vorticity} toroidally averaged (b), and with the interchange drive term $C(p_e+\tau p_i)$ in Eq.~\eqref{eqn:vorticity} toroidally averaged (c).}
    \label{fig:no_kh}
\end{figure}

Fig.~\ref{fig:no_kh} displays a typical snapshot of the electron temperature for a simulation in the suppressed transport regime ($s_{T0}=0.6$ and $\nu_0=0.9$) and compares it to two simulations having the same parameters, but with the KH and ballooning drives removed. 
Turbulence is strongly suppressed when the Reynolds stress is toroidally averaged. 
On the other hand, no significant effect on turbulence is noticed when the interchange drive is removed.
This shows that the KH instability is the main drive of turbulent transport and, consequently, regulates the equilibrium pressure gradient in the suppressed transport regime. 

An analytical estimate of the equilibrium pressure gradient length in the edge  when turbulence is driven by the KH instability can be derived by following a procedure similar to the one detailed in \S\ref{sec:interchange} and discussed for a linear device by \citet{rogers2010}. 
The growth rate of the KH instability is proportional to the $\mathbf{E}\times\mathbf{B}$ shear \citep{Myra2016}, $\gamma_\text{\tiny{KH}} \sim \rho_*^{-1}\partial_{rr}^2 \bar{\phi} \sim \rho_*^{-1}\bar{T}_e/L_p^2$, having assumed $L_\phi \sim L_p$. Being KH a global mode, the size of the turbulent eddies it generates is comparable to the pressure gradient length, $k_{\psi,\text{\tiny{KH}}} \sim 1/L_p$. 
Therefore, similarly to Eq.~\eqref{eqn:transport}, the KH-driven heat flux can be expressed as 
\begin{equation}
    \label{eqn:transport_kh}
    q_{\psi,\text{\tiny{KH}}} \sim \frac{\bar{T}_e \bar{p}_e}{L_p}\,.
\end{equation}
By balancing the heat source integrated over the region inside the LCFS and the perpendicular turbulent heat flux crossing the LCFS, similarly to Eq.~\eqref{eqn:flux_balance}, but assuming that $q_{\psi,\text{\tiny{KH}}}$ is approximately uniform along the LCFS, we obtain
\begin{equation}
    \label{eqn:lp_kh}
    L_{p,\text{\tiny{KH}}} \sim \frac{\bar{p}_e \bar{T}_e}{4S_p}L_\chi\,,
\end{equation}
where $\bar{T}_e$ and $\bar{p}_e$ are evaluated at the LCFS.

\begin{figure}
    \centering
    \includegraphics[scale=0.65]{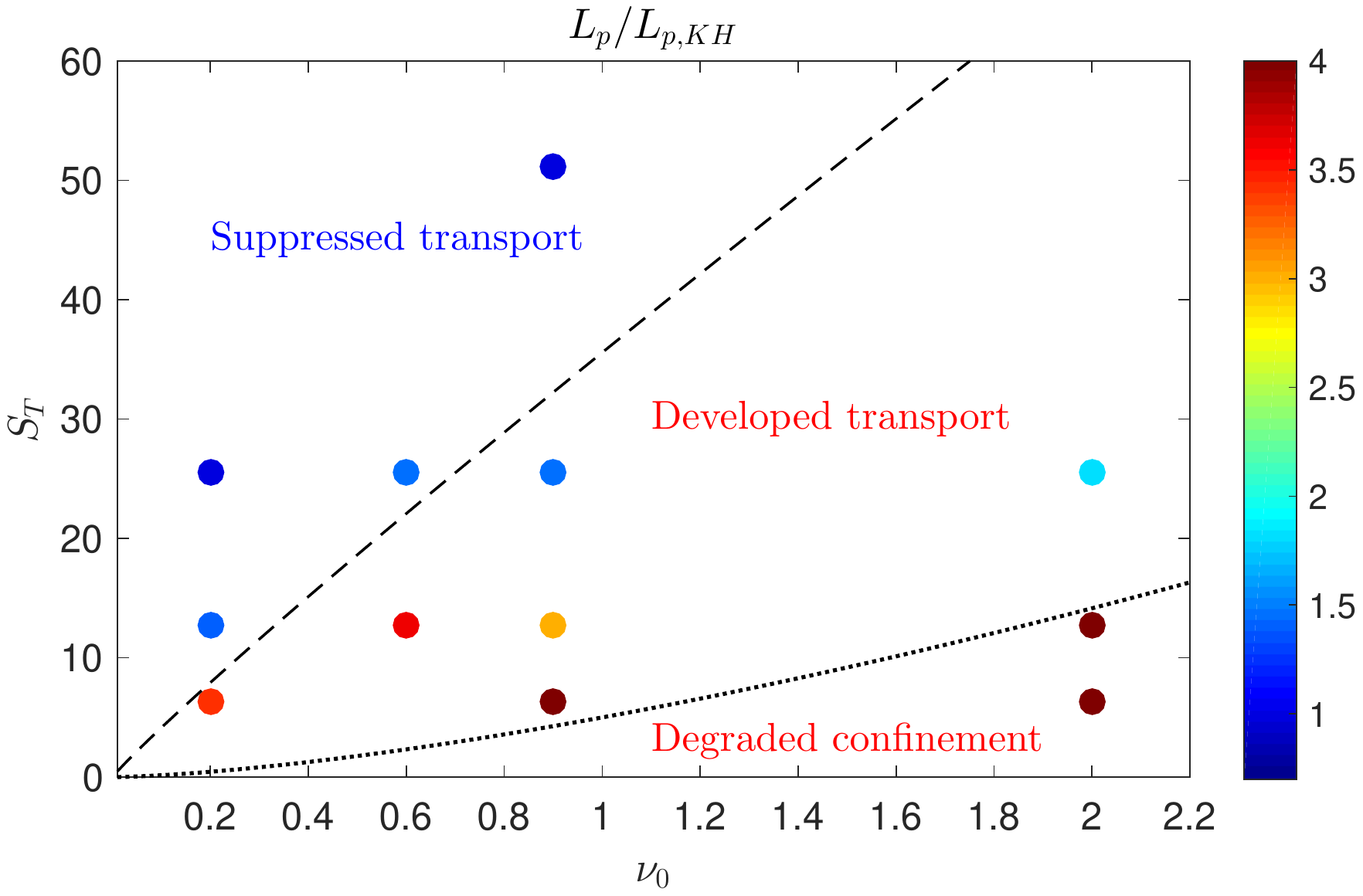}
    \caption{Ratio of $L_p$, the equilibrium pressure gradient length directly obtained from the simulations, to $L_{p,\text{\tiny{KH}}}$, the estimate in Eq.~\eqref{eqn:lp_kh} based on the assumption that the transport is driven by the KH instability, for all the simulations considered in the present study. The dashed black line represents the heat source threshold to access the suppressed transport regime, Eq.~\eqref{eqn:temp_source}, while the dotted black line the threshold to access the degraded confinement regime, Eq.~\eqref{eqn:temperature_den_limit}.}
    \label{fig:lp_kh}
\end{figure}

The ratio of $L_p$, the equilibrium pressure gradient length directly obtained from the simulations, to $L_{p,\text{\tiny{KH}}}$ is displayed in Fig.~\ref{fig:lp_kh} for the different simulations considered in the present study. At large values of $S_T$ and small $\nu_0$, a region can be identified where $L_{p,\text{\tiny{KH}}}$ well reproduces the simulation results. In fact, the results of Figs.~\ref{fig:lp_int}~and~\ref{fig:lp_kh} show that turbulent transport is driven by the KH instability in the suppressed transport regime, otherwise the interchange instability regulates the equilibrium pressure gradient length.
Furthermore, we note that $L_p>L_{p,\text{\tiny{KH}}}$ in the ballooning-driven parameter region, while $L_p>L_{p,i}$ in the suppressed transport regime, as expected from the fact that the mode driving turbulence minimises the pressure gradient. 

The suppressed transport regime shows some of the main key aspects observed experimentally in H-mode discharges, such as the presence of a strong sheared flow (see Fig.~\ref{fig:profiles} (c)), the reduction of the turbulence level with respect to the L-mode that leads to the formation of a transport barrier near the separatrix (see Fig.~\ref{fig:profiles} (a)), and the increase of the energy confinement time (see Fig.~\ref{fig:conf_time}). All this occurs when a power threshold is exceeded, as detailed in \S\ref{sec:power_threshold}. We therefore associate the suppressed transport regime to the H-mode of tokamak operation.

\subsection{Degraded confinement regime}\label{sec:density_limit}

\begin{figure}
    \centering
    \includegraphics[scale=0.65]{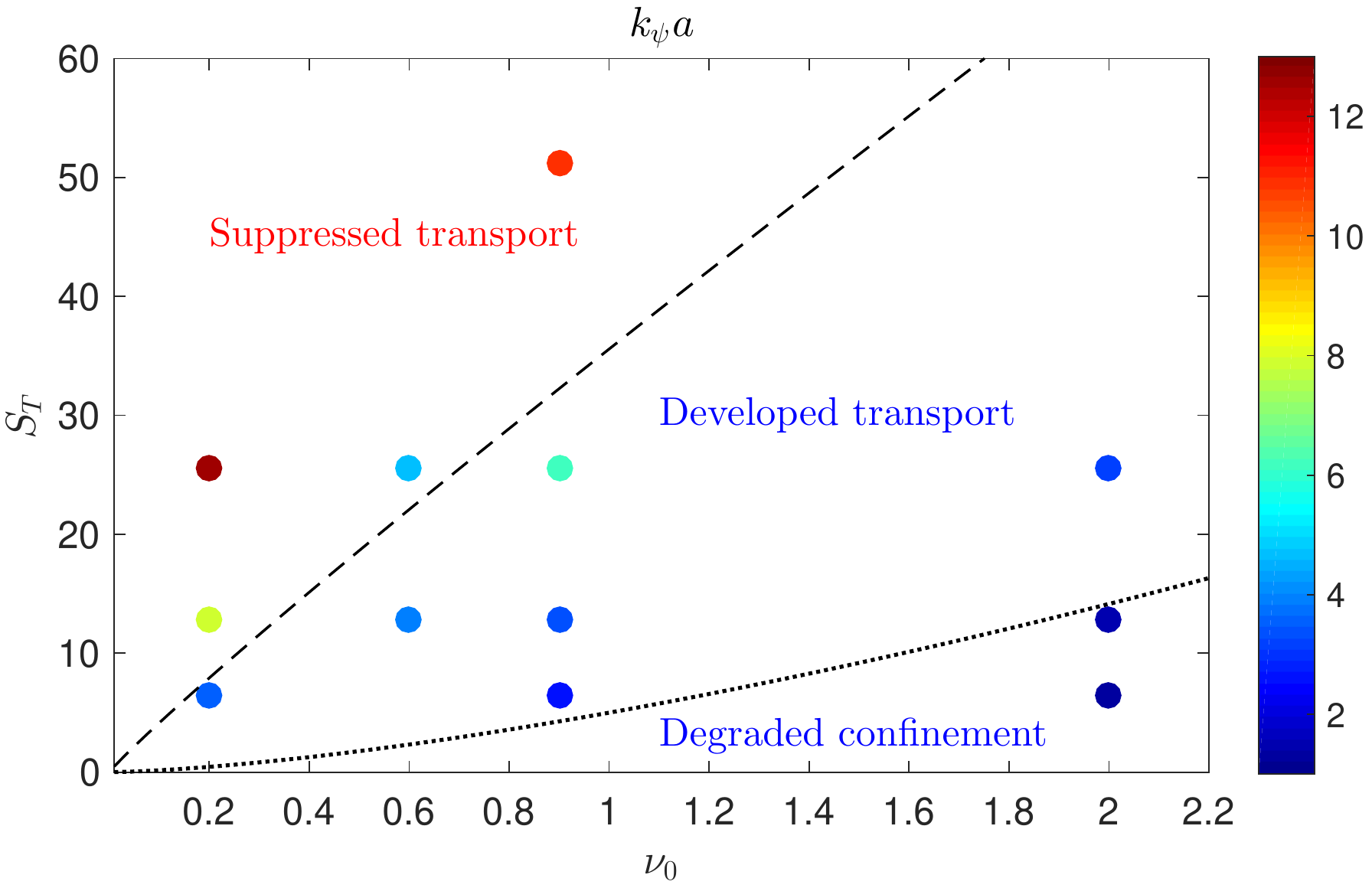}
    \caption{Radial wavenumber computed from the correlation length of turbulent eddies, normalized to the tokamak minor radius, for all the simulations considered in the present study. The dashed black line represents the heat source threshold to access the suppressed transport regime, Eq.~\eqref{eqn:temp_source}, while the dotted black line the threshold to access the degraded confinement regime, Eq.~\eqref{eqn:temperature_den_limit}.}
    \label{fig:radial_extension}
\end{figure}

Figs.~\ref{fig:den_limit}~and~\ref{fig:conf_time} show that plasma turbulence and confinement properties strongly vary also within the interchange driven turbulent regime.
In general, for high values of $\nu_0$ and low values of $s_{T0}$, poor confinement properties and a catastrophically large turbulent transport are observed. Indeed, in this parameter regime, despite being described as the non-linear development of a ballooning mode, turbulence results into high level fluctuations, with amplitude comparable to the equilibrium quantity, that propagate from the edge to the core region, as shown in Fig.~\ref{fig:den_limit}. This is due to the fact that the radial size of the turbulent structures increases with $\nu_0$, since $k_{\chi,i} \propto \nu_0^{-1/2}$ and $k_{\psi,i} \simeq \sqrt{k_{\chi,i}/L_{p,i}} \propto \nu_0^{-7/12}$.
Fig.~\ref{fig:radial_extension} shows the radial extension of turbulent eddies normalized to the tokamak minor radius for the different simulations considered in the present study. 
In particular, if the value of $\nu_0$ is sufficiently large and $s_{T0}$ sufficiently small, turbulent eddies appear to have a size comparable to the tokamak minor radius, i.e. $k_\psi a\sim 1$. As a consequence, they extend towards the core region and lead to a very large cross-field turbulent transport throughout the closed field line region.  
In these conditions, the core temperature can significantly drop and MHD modes, which are beyond the description provided by our model, can play an important role \citep{greenwald1988,greenwald2002}.
As already mentioned in~\S\ref{sec:overview}, the degraded confinement regime is linked to high values of the density and we associate it to the crossing of the density limit, in agreement with the result of~\citet{hajjar2018}.  Experimental evidences that the density limit is due to an increase of edge collisionality, proportional to $\nu_0$ in the model considered, are reported from the TJ-K stellarator \citep{schmid2017}.

\begin{figure}
    \centering
    \includegraphics[scale=0.7]{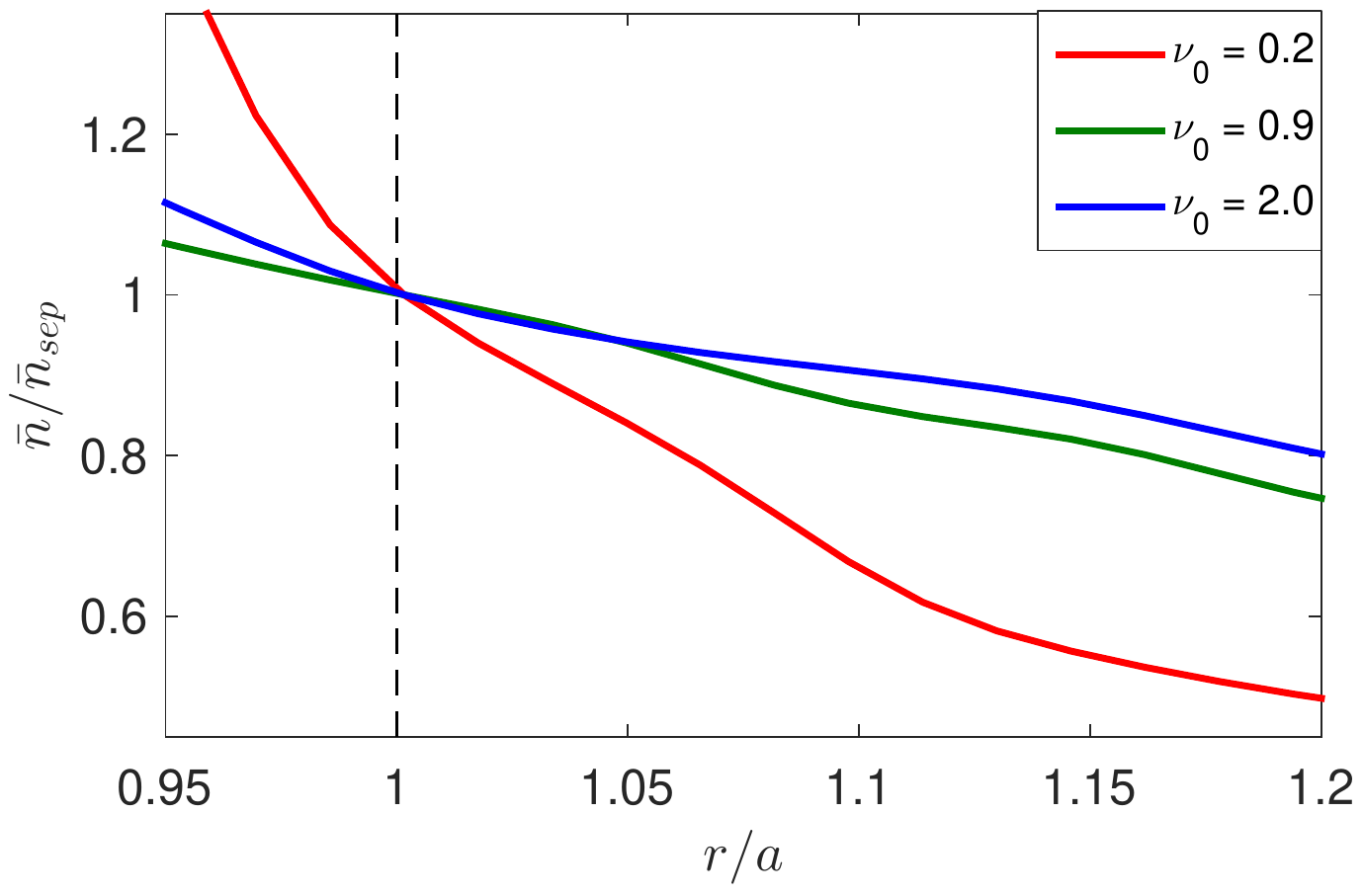}
    \caption{Equilibrium density radial profile at the outboard midplane for simulations at $s_{T0}=0.075$ and different values of $\nu_0$.}
    \label{fig:den_prof}
\end{figure}

Three main effects are observed when crossing the density limit. First, the $\mathbf{E}\times\mathbf{B}$ shear near the separatrix in the degraded confinement regime is even weaker than in the developed transport regime, as shown in Fig.~\ref{fig:profiles}~(c). This is in agreement with recent experiments that show how the edge shear flow collapses when the density limit is approached \citep{hong2017}.
Therefore, the $\mathbf{E}\times\mathbf{B}$ shear is an important quantity not only to explain the transition from the developed transport regime to the suppressed transport regime, but also to recognize the crossing of the density limit.
Second, the degraded confinement regime is characterized by a flatter equilibrium density profile in the SOL with respect to the developed and suppressed transport regimes (see Fig.~\ref{fig:den_prof}). In fact, the blob size increases with the collisionality \citep{ippolito2011,nespoli2017,beadle2020}, leading to an enhancement of the cross-field turbulent transport in the far SOL. The density profile becomes flatter with no clear distinction between the edge, near SOL and far SOL. Experimental observations of the flattening of the density profiles as the density increases towards the density limit are reported by~\citet{labombard2001}.  
Third, the large amplitude fluctuations that extend towards the core region lead to a strong enhancement of cross-field turbulent transport and the loss of confinement.

\section{Transition threshold between transport regimes}\label{sec:transitions}

In this section, we focus on the transition from the developed transport regime to the suppressed transport regime, which we associate to the L-H transition, and from the developed transport regime to the degraded confinement regime, which we associate to the crossing of the density limit. Analytical estimates of the heat source threshold to access the suppressed transport regime and of the density threshold to access the degraded confinement regime are derived.

\subsection{Heat source threshold to access the suppressed transport regime\label{sec:power_threshold}}

The transition from the L-mode to the H-mode occurs when $L_{p,i}\simeq L_{p,{\text{\tiny{KH}}}}$, namely when the turbulent transport due to the interchange instability equals the one due to the KH instability.
An estimate of $S_p$ at the transition can be derived by equating Eqs.~\eqref{eqn:lp_int}~and~\eqref{eqn:lp_kh}, 
\begin{equation}
    \label{eqn:power_source_intermediate}
    S_p^\text{\tiny{LH}}\sim \rho_* L_\chi \nu_0^2 q_{95}^4 \bar{n}^3/(2 \bar{T}_e)\,,
\end{equation}
where $\bar{n}$ and $\bar{T}_e$ are evaluated at the LCFS.
The relation between $\bar{T}_e$ at the LCFS and $S_p$ can be obtained by balancing $S_p$ with the parallel losses to the vessel walls. As an order of magnitude estimate, this balance can be expressed by using the integral of the heat flux over the SOL width, $\Delta_{\text{SOL}}$, as
\begin{equation}
    \label{eqn:sol_balance_first}
    \int_{\Delta_{\text{SOL}}} \bar{p}_e \bar{c}_s\, \mathrm{d}l \sim S_p\,,
\end{equation}
having assumed of being in the sheath connected regime (i.e. no temperature drop in the divertor region) and assuming that the plasma outflows at the divertor plate with the sound speed. 
Furthermore, by assuming that the pressure and temperature decay in the SOL exponentially on the $L_p$ scale, Eq.~\eqref{eqn:sol_balance_first} becomes
\begin{equation}
    \label{eqn:sol_balance}
    \bar{T}_e \sim \biggl(\frac{2 S_p}{\bar{n} L_p}\biggr)^{2/3}\,,
\end{equation}
with $\bar{T}_e$ and $\bar{n}$ evaluated at the LCFS.
Eqs.~\eqref{eqn:power_source_intermediate}~and~\eqref{eqn:sol_balance} allow us to derive an analytical estimate of the heat source threshold for H-mode access,
\begin{equation}
    \label{eqn:power_threshold}
    S_p^\text{\tiny{LH}} \sim \rho_*^{7/15}(\nu_0 q_{95}^2)^{14/15} L_\chi^{11/15}\bar{n}^{29/15}\,,
\end{equation}
the corresponding electron temperature at the LCFS,
\begin{equation}
    T_e^\text{\tiny{LH}}\sim \rho_*^{4/15} (\nu_0 q_{95}^2)^{8/15} L_\chi^{2/15} \bar{n}^{8/15} ,
\end{equation}
and the equilibrium pressure gradient length at the transition,
\begin{equation}
    L_p^\text{\tiny{LH}}\sim \rho_*^{1/15} (\nu_0 q_{95}^2)^{2/15} L_\chi^{8/15} \bar{n}^{2/15}\ .
\end{equation}
In Eq.~\eqref{eqn:power_threshold}, the increase of the heat source required to access the H-mode with $\nu_0$ is due to the increase of cross-field turbulent transport in the developed transport regime. Indeed, $q_{\psi,i}$ is proportional to $\nu_0^{1/2}$ (see Eq.~\eqref{eqn:transport_int}) and $L_{p,i}$ is proportional to $\nu_0^{2/3}$ (see Eq.~\eqref{eqn:lp_int}). 

We now compare our analytical estimate with the simulation results. For this purpose, we express Eq.~\eqref{eqn:power_threshold} in terms of $S_T$, by using $S_T^\text{\tiny{LH}}\simeq S_p^\text{\tiny{LH}}/\bar{n}$, and we obtain  
\begin{equation}
    \label{eqn:temp_source}
    S_T^\text{\tiny{LH}} \sim \rho_*^{7/15}(\nu_0 q_{95}^2)^{14/15} L_\chi^{11/15}\bar{n}^{14/15}\,.
\end{equation}
The analytical estimate of the threshold $S_T^\text{\tiny{LH}}$ as a function of $\nu_0$ (assuming a constant value for the normalized density $\bar{n}$ at the LCFS) is displayed in Fig.~\ref{fig:lp_kh}, showing a very good agreement between the analytical prediction of Eq.~\eqref{eqn:temp_source} and the simulation results.

We also link our L-H transition with experimental observations.
In order to identify the scaling of $S_p^\text{\tiny{LH}}$ with the main experimental parameters, we write the power threshold in Eq.~\eqref{eqn:power_threshold} in physical units
\begin{equation}
    \label{eqn:power_physical}
    \begin{split}
    P_{\text{\tiny{LH}}} = 2\pi R_0 S_p^\text{\tiny{LH}} &\simeq 9\times 10^7\Bigl(\frac{m_e}{m_i}\Bigr)^{9/15} n^{29/15} R_0^{22/15} q_{95}^{28/15} a^{11/15} B_T^{-11/15}\\
    &\simeq 9\times 10^7\Bigl(\frac{m_e}{m_i}\Bigr)^{0.6}n^{1.9} R_0^{1.5} q_{95}^{1.9}a^{0.7}B_T^{-0.7},
    \end{split}
\end{equation}
with the $2\pi R_0$ factor taking into account the integration of the heat source along the toroidal direction, having imposed $L_\chi\sim 2\pi a$, and the density at the LCFS being expressed in units 10$^{20}$~m$^{-3}$. 
The scaling law in Eq.~\eqref{eqn:power_physical} correctly reproduces the isotope effect observed in the experiments \citep{righi1999,maggi2017} and also found in previous theoretical investigations \citep{dominici2019}.  The dependence on $a$ and $R_0$ shows a good agreement with the experimental scaling law in Eq.~\eqref{eqn:power_experiments}. The exponent of the density in Eq.~\eqref{eqn:power_physical} is approximately a factor 2.7 larger than the one predicted by the experimental scaling law in Eq.~\eqref{eqn:power_experiments}, although we remark that the density in Eq.~\eqref{eqn:power_physical} is evaluated at the LCFS, while the density in Eq.~\eqref{eqn:power_experiments} denotes the line-averaged density. 
The power threshold in Eq.~\eqref{eqn:power_physical} depends inversely on the toroidal magnetic field, while the experimental scaling law in Eq.~\eqref{eqn:power_experiments} shows a direct dependence on $B_T$. Moreover, in contrast to the experimental scaling law in Eq.~\eqref{eqn:power_experiments}, the power threshold in Eq.~\eqref{eqn:power_physical} depends on $q_{95}$. 

As an example of the evaluation of Eq.~\eqref{eqn:power_physical} in experimental conditions, we consider the value of the power threshold predicted for typical parameters of the TCV tokamak ($a=0.25$~m, $R_0=0.88$~m, line-averaged density $n_e\simeq 4\times 10^{19}$~m$^{-3}$, density at the LCFS $n\simeq 2\times 10^{19}$~m$^{-3}$, $B_T\simeq 1.4$~T, and $q_{95} \simeq 4$). 
The estimate in Eq.~\eqref{eqn:power_physical} gives $P_{\text{\tiny{LH}}}\simeq$~142~kW, a power threshold that has the same order of magnitude as the experimental TCV power threshold, $P_{\text{\tiny{LH}}}\simeq 260$~kW (see, e.g., \citet{scaggion2012,martin2014}). 

Experimental measurements show that the power to access the H-mode is lower when the ion-$\nabla B$ drift direction is towards the X-point, rather than away from it \citep{asdex1989}. In order to study the dependence of the heat source threshold on the ion-$\nabla B$ drift direction, we consider two simulations  where we vary the direction of the toroidal magnetic field while keeping the same direction of the plasma current and other parameters the same. In particular, we consider $s_{T0}=0.3$ and $\nu_0=0.9$, parameters close to the L-H transition when the ion-$\nabla B$ drift direction points upwards (unfavorable for H-mode access).  
The equilibrium density, temperature and $\mathbf{E}\times \mathbf{B}$ shear profiles at the LFS midplane do not show significant differences between the simulations with favorable and unfavorable ion-$\nabla B$ drift direction, both simulations belonging to the developed transport regime. 
Therefore, at least in the case analysed, the power threshold to access the H-mode is independent of the toroidal magnetic field direction in our model.
The discrepancy between experimental and simulation observations may be due to the absence of kinetic effects involving passing and trapped particles, among these we mention the effects of ion orbit loss, which can be important in establishing the dependence of the L-H transition on the ion-$\nabla B$ drift direction \citep{stoltzfus2012,boedo2016}, as also pointed out by XGC1 simulations \citep{ku2018} .

\begin{figure}
    \centering
    \includegraphics[scale=0.62]{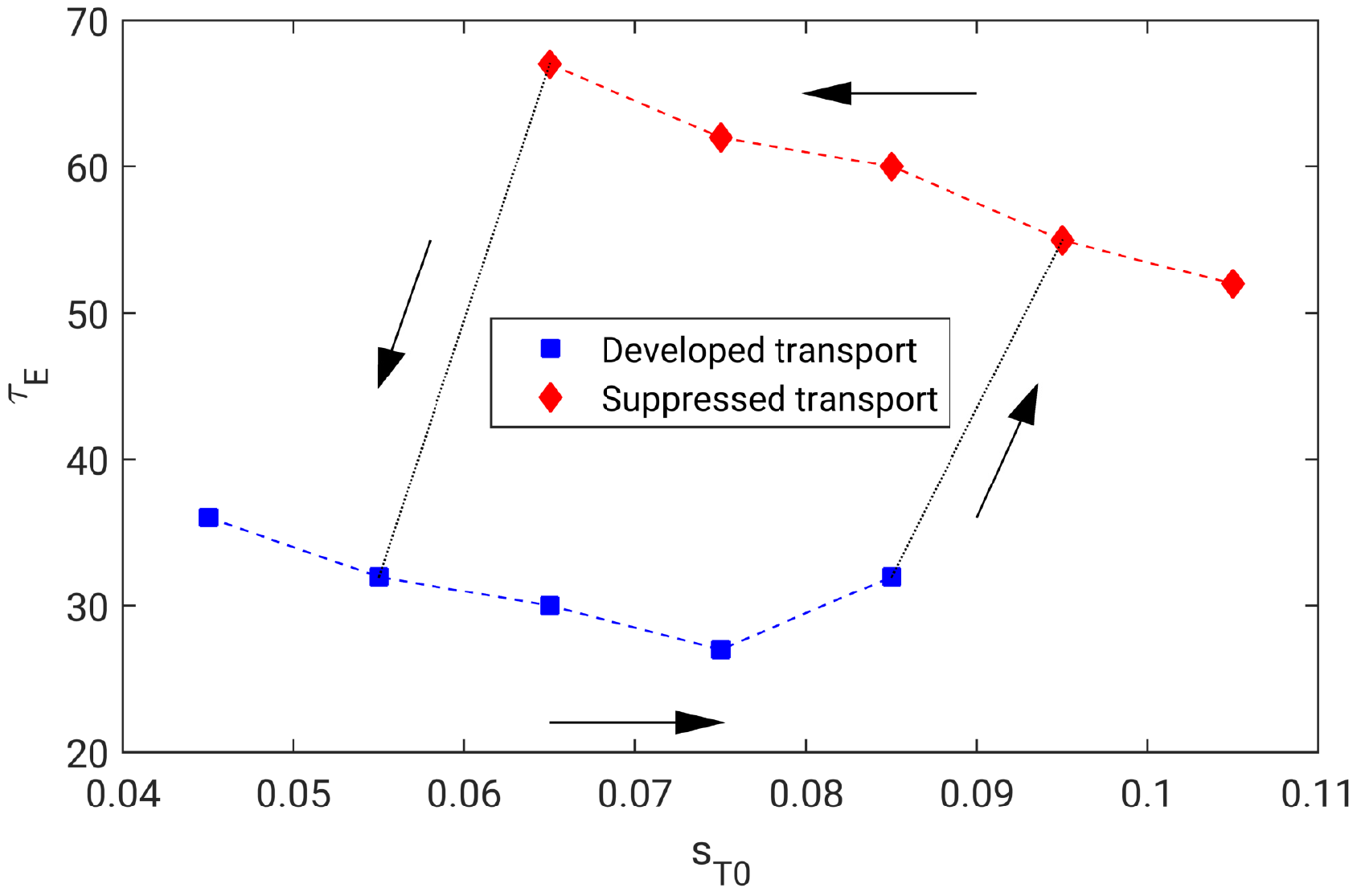}
    \caption{Energy confinement time for simulations with $\nu_0=0.2$ and different values of $s_{T0}$. Simulations in the developed transport regime are denoted by blue square markers, in the suppressed transport regime by red diamond markers. Starting from a simulation in the developed transport regime, $s_{T0}$ is progressively increased from 0.045 to 0.105. The transition to the suppressed transport regime occurs approximately at $s_{T0}\simeq 0.085$. The heat source is then progressively reduced until the reverse transition occurs, approximately at $s_{T0}\simeq 0.065$. The transitions are represented as dotted black line.}
    \label{fig:hysteresis}
\end{figure}

Experimental observations \citep{Snipes1996,Thomas1998} and theoretical models \citep{Drake1996,Hinton1991} point out the presence of hysteresis on the power threshold for the L-H transition, i.e. having entered the H-mode conditions, the hysteresis allows for a decrease of the power below the threshold for H-mode access without inducing the H-L transition. 
The presence of hysteresis in our simulations is investigated by performing a set of simulations at $\nu_0=0.2$ and different values of $s_{T0}$ in the proximity of the threshold to access the suppressed transport regime (more precisely we consider $s_{T0}=$ 0.045, 0.055, 0.065, 0.075, 0.085, 0.095, 0.105).
Starting from a simulation in the developed transport regime, $s_{T0}$ is progressively increased from 0.045 to 0.105 where the transition to the suppressed transport regime occurs. Then, by using the simulation at $s_{T0}=0.105$ in the suppressed transport regime as initial condition, we perform a second set of simulations where $s_{T0}$ is progressively reduced, observing the H-L transition at $s_{T0} \simeq 0.065$ (see Fig.~\ref{fig:hysteresis}). Therefore, the transition from the developed transport regime to the suppressed transport regime occurs at a higher value of the heat source than the reverse transition, thus pointing out the presence of hysteresis in the considered model. 

The presence of hysteresis can be explained as follows. In the suppressed transport regime, the $\mathbf{E}\times\mathbf{B}$ shear is strong near the separatrix and the turbulent transport is mainly driven by the KH instability. As the heat source decreases, the equilibrium pressure gradient decreases (see Eq.~\eqref{eqn:lp_kh}) as well as the $\mathbf{E}\times\mathbf{B}$ shear near the separatrix. 
However, the $\mathbf{E}\times\mathbf{B}$ shear remains sufficiently strong to stabilize ballooning modes, thus allowing for a decrease of the heat source below the L-H transition threshold with no collapse of the $\mathbf{E}\times\mathbf{B}$ shear. This collapse is suddenly followed by the onset of the interchange instability, with the developed transport regime eventually reached.  

As an aside observation of Fig.~\ref{fig:hysteresis}, we note that, within the same transport regime, the energy confinement time decreases as the heat source increases, the only exception being the simulation at $s_{T0}=0.085$ in the developed transport regime, which is in proximity of the transition. The decrease of the energy confinement time following the increase of the heat source is also observed in many experiments \citep{yushmanov1990,cordey2005}.

\subsection{Density threshold to access the degraded confinement regime}
\label{sec:density_threshold}

In our simulations, the transition to the degraded confinement regime occurs gradually as the edge fluctuations present in the developed transport regime reach a size comparable to the system size, $1/k_\psi\sim a$, and the equilibrium pressure gradient length becomes comparable to the tokamak minor radius, $L_p\sim a$. This last observation can be considered as a condition to access the regime of degraded confinement. By assuming interchange-driven turbulent transport, this condition can be expressed as
\begin{equation}
    \label{eqn:cond_den_lim}
    a\sim L_{p,i}\sim \biggl[\frac{\rho_*}{2}(\nu q_{95}^2 \bar{n})^2\biggl(\frac{2\pi a}{S_p}\bar{p}_e\biggr)^4\biggr]^{1/3}\bar{T}_e,
\end{equation}
having estimated the equilibrium pressure gradient length according to Eq.~\eqref{eqn:lp_int}, and $\bar{n}$ and $\bar{T}_e$ being evaluated at the LCFS. By using the estimate of the electron temperature in Eq.~\eqref{eqn:sol_balance}, we obtain
\begin{equation}
    \label{eqn:power_den_limit}
    S_p\sim  \rho_*^{3/4}\frac{8\pi^3}{a^{5/4}}(\nu_0 q_{95}^2)^{3/2}\bar{n}^{5/2}\,. 
\end{equation}
In order to compare Eq.~\eqref{eqn:power_den_limit} with the simulation results, we express Eq.~\eqref{eqn:power_den_limit} in terms of $S_T$, again using $S_p\simeq S_T/\bar{n}$, and we find 
\begin{equation}
    \label{eqn:temperature_den_limit}
    S_T \sim \rho_*^{3/4}\frac{8\pi^3}{a^{5/4}}(\nu_0 q_{95}^2)^{3/2}\bar{n}^{3/2}\,.
\end{equation}
The analytical prediction of temperature source threshold in Eq.~\eqref{eqn:temperature_den_limit} is plotted in the phase space of Fig.~\ref{fig:radial_extension}, by keeping constant the normalized density at the LCFS (this value is approximately constant in the simulations considered in the present study). A good agreement between the simulation results and the theoretical estimate is shown by the analysis of the radial extension of the turbulent eddies.

For comparison with experiment results, the heat source threshold in Eq.~\eqref{eqn:power_den_limit} can be written in physical units,
\begin{equation}
    \label{eqn:power_den_phys}
    S_p\simeq 3.2 \Bigl(\frac{m_i}{m_e}\Bigr)^{1/4} \frac{a^{19/4} n^{5/2}}{I_p^3 R_0^{9/4}}\,,
\end{equation}
where $S_p$ is expressed in kW/m and $n$ in 10$^{20}$~m$^{-3}$, $I_p$ is the plasma current (in MA), and we have used $q_{95}\sim 2\pi a^2 B_T/(R_0 I_p)$.   
The density threshold to access the degraded confinement regime, corresponding to the operational density limit evaluated at the LCFS, can then be derived from Eq.~\eqref{eqn:power_den_phys} and, in physical units, takes the following form
\begin{equation}
    \label{eqn:den_limit}
    n \simeq 0.3 \Bigl(\frac{m_e}{m_i}\Bigl)^{1/10}  P_\text{sep}^{2/5} R_0^{1/2} \frac{I_p^{6/5}}{a^{19/10}} \simeq 0.3 \Bigl(\frac{m_e}{m_i}\Bigl)^{0.1}  P_\text{sep}^{0.4} R_0^{0.5} \frac{I_p^{1.2}}{a^{1.9}}\,, 
\end{equation}
where $P_\text{sep}=2\pi R_0 S_p$ is the power crossing the separatrix (in kW), with $I_p$ in MA and $n$ in 10$^{20}$~m$^{-3}$.
The comparison between the analytical scaling law of Eq.~\eqref{eqn:den_limit} and the empirical scaling in Eq.~\eqref{eqn:greenwald} is not straightforward since we have expressed the density limit in terms of the density at the LCFS, while the empirical scaling refer to the line-averaged density. With this caveat in mind, we note that Eq.~\eqref{eqn:den_limit} reproduces the dependence on the plasma current and tokamak minor radius expected by the experimental scaling law of Eq~\eqref{eqn:greenwald}. The dependence on the ion mass is weak, in agreement with experimental observations that do not show evident isotope effect on the density limit \citep{saibene1999}. The analytical scaling law of Eq.~\eqref{eqn:den_limit} depends on the heat source, a feature also observed in some experiments \citep{stabler1992,mertens1997}. For instance, it has been experimentally found \citep{mertens1997} that, at the density limit, the density at the LCFS depends on the power crossing the separatrix as $n \propto P_\text{sep}^{0.6}$, in good agreement with the power dependence in Eq.~\eqref{eqn:den_limit}. However, the analytical density threshold of Eq.~\eqref{eqn:den_limit} depends also on the tokamak major radius, while the empirical scaling in Eq.~\eqref{eqn:greenwald} is independent of it.
As an example of application of Eq.~\eqref{eqn:den_limit}, we note that the threshold density predicted for the TCV tokamak ($a=0.25$~m, $R_0=0.88$~m, $I_p=1$~MA, and $P_\text{sep}=100$~kW) is approximately $n \simeq 10^{21}$~m$^{-3}$.

\section{Conclusions\label{sec:conclusions}}

In the present paper, results of flux-driven simulations in realistic single-null geometry, carried out by using the GBS code with the domain encompassing the whole tokamak to retain the core-edge-SOL interplay, are used to study the important role of sources and resistivity in driving a variety of turbulent transport regimes in the tokamak edge. Our simulations show the presence of three turbulent transport regimes:  a regime of developed turbulent transport, which we link to the L-mode observed in the experiments, a regime of suppressed turbulent transport, with similarities to the H-mode, and a regime of degraded confinement, which we associate to the crossing of the density limit.
The developed transport and degraded confinement regimes appear at low heat source and high resistivity, with turbulent transport driven by the interchange instability, while the suppressed transport regime appears at high heat source and low resistivity, with turbulent transport driven by the KH instability. 
The energy confinement time in the suppressed transport regime is approximately a factor of two higher than in the developed transport regime. An overall loss of confinement is observed in the degraded confinement regime, with strong fluctuations that reach the tokamak core.
An analytical expression of the equilibrium pressure gradient length in the tokamak edge is derived for all the transport regimes.

The transition from the developed to the suppressed transport regime shows many features in common with the L-H transition observed experimentally, such as the presence of a strong sheared flow, the reduction of the turbulence level, the formation of a transport barrier near the separatrix and the presence of a power threshold.  
The power threshold for H-mode access derived herein is able to reproduce the isotope effect and the experimental parameters scaling of Eq.~\eqref{eqn:power_experiments}, with the exception of the toroidal magnetic field and the dependence on safety factor. In addition, no dependence of the power threshold on the ion-$\nabla B$ direction is observed in the considered simulations. 
The transition from the developed to the suppressed transport regime is subject to hysteresis as it occurs at a higher value of the heat source with respect to the inverse transition.  
The analytical prediction of the power threshold shows a good agreement with the results of GBS simulations performed with different values of heat source and resistivity (see Fig.~\ref{fig:conf_time}).   

In the degraded confinement regime, found at high values of resistivity and low heat source, turbulent transport is driven by the interchange instability with turbulent eddies of size comparable to the tokamak minor radius. High-level fluctuations are generated in the core and the particle confinement time drops. 
We derive an analytical estimate of the density threshold to access the degraded confinement regime, which we associate to the operational Greenwald density limit. Indeed, it retrieves the main dependencies on the plasma current and tokamak minor radius observed experimentally \citep{greenwald2002}.

Finally, we remark that the model considered in this work neglects coupling with neutrals dynamics, neoclassical and kinetic effects. Moreover, it is electrostatic and makes use of the Boussinesq approximation. 
These terms can definitely have an impact on the edge turbulent regimes. In fact, neutral dynamics may affect the L-H transition dynamics, as shown by~\citet{shaing1995,carreras1996,owen1998}.
In~\citet{chone2014,chone2015,viezzer2013}, it is shown that neoclassical terms play an important role on the radial electric field responsible for the onset of a transport barrier that leads to the L-H transition. Kinetic effects can also be important \citep{stoltzfus2012,boedo2016}.
Electromagnetic effects can play a role in setting the pedestal height and width \citep{snyder2004,snyder2009}, and the density limit \citep{rogers1998}. 
In~\citet{bodi2011,stegmeir2019}, the validity of the Boussinesq approximation is addressed and, while the results show that a good agreement exists between the results of turbulence simulations that make use or avoid the application of the Boussinesq approximation, this result cannot be taken for granted in general.
As a future work, we plan to include these effects for a more accurate investigation of the edge turbulent regimes and to study separately the role of the ion and electron temperature source on the transitions. 

\section*{Acknowledgments}
The authors thank D. Galassi, Y. Martin, and C. Theiler for useful discussions.
The simulations presented herein were carried out in part at the Swiss National Supercomputing Center (CSCS) under the project ID s882 and in part on the CINECA Marconi supercomputer under the GBSedge project. 
This work, supported in part by the Swiss National Science Foundation, was carried out within the framework of the EUROfusion Consortium and has received funding from the Euratom research and training programme 2014 - 2018 and 2019 - 2020 under grant agreement No 633053. The views and opinions expressed herein do not necessarily reflect those of the European Commission.

\bibliographystyle{jpp}

\bibliography{bibliography}

\end{document}